\documentclass[reprint,superscriptaddress,amsmath,amssymb,aps,pre]{revtex4-2}
\usepackage{graphicx}
\usepackage{dcolumn}
\usepackage{bm}
\usepackage{float}
\usepackage{times}
\usepackage[unicode=true,pdfusetitle,bookmarks=false,colorlinks=true,citecolor=blue,urlcolor=blue,linkcolor=blue]{hyperref}

\begin{document}
		
	\title{Active Particles Destabilize Passive Membranes}
	
	\author{David A. King}
	\email{daking2@lbl.gov}
	\affiliation{Materials Science Division, Lawrence Berkeley National Laboratory, Berkeley CA 94720.}
	\affiliation{Department of Materials Science and Engineering, University of California, Berkeley, CA 94720.}
    
    \author{Thomas P. Russell}
    \affiliation{Polymer Science and Engineering Department, University of Massachusetts, Amherst MA 01003.}
    \affiliation{Materials Science Division, Lawrence Berkeley National Laboratory, Berkeley CA 94720.}
    
	\author{Ahmad K. Omar}
    \email{aomar@berkeley.edu}
	\affiliation{Materials Science Division, Lawrence Berkeley National Laboratory, Berkeley CA 94720.}
	\affiliation{Department of Materials Science and Engineering, University of California, Berkeley, CA 94720.}
    	
	\begin{abstract}
	We present a theory for the interaction between active particles and a passive flexible membrane. 
    By explicitly solving for the pressure exerted by the active particles, we show that they \textit{reduce} the membrane tension and bending modulus and introduce novel \textit{non-local} contributions to the membrane mechanics.  
    This theory predicts activity-induced instabilities and their morphology are in agreement with recent experimental and simulation data.
	\end{abstract}
	
	\maketitle
	
Cells are intrinsically active systems, maintained far from equilibrium by continuous energy consumption. 
Two hallmark behaviors, reproduction by mitosis and feeding by phagocytosis, require major morphological manipulations mediated by active cytoskeletal forces on the cell membrane~\cite{May2001,Pollard2009,Rottner2019,Welf2020}.
Pathogenic bacteria such as \textit{Listeria} and \textit{Shigella} similarly harness localized active forces to deform host cell membranes and invade neighboring cells~\cite{Pizarro-Cerd2016,Friedrich2012}.
Understanding how localized active stresses induce large-scale deformations in flexible membranes is key to unraveling these processes.

Recently, synthetic model systems, such as vesicles loaded with bacteria~\cite{Takatori2020}, Janus colloids~\cite{Vutukuri2020} or magnetic rollers~\cite{Kim2025}, have allowed the interplay between active mechanics and membrane response to be examined.  
Through interactions with the active particles, the vesicles can exhibit non-equilibrium fluctuation spectra~\cite{Takatori2020} and extreme shape instabilities~\cite{Vutukuri2020}, including division~\cite{Kim2025}.
	
In this Letter, we present a theoretical framework to describe these instabilities. 
As a canonical example, we consider hard, spherical active Brownian particles (ABPs) with radius $b$ confined to one side of an infinite membrane. 
This geometry is appropriate for exceedingly large vesicles and enables analytical progress while capturing the essential physics. 
Similarly, we consider ``dry'' ABPs with no hydrodynamics to compare with recent simulations~\cite{Vutukuri2020} performed in the same limit.
By explicitly solving for the pressure exerted by ABPs on the fluctuating membrane, we show that their influence can be captured by renormalizing the membrane’s physical properties: they introduce \textit{non-local interactions} and can \textit{reduce} both the surface tension and the bending modulus. 
Crucially, when these effective parameters become negative, a membrane \textit{stable} in equilibrium can undergo an \textit{activity-induced instability}, the qualitative thresholds and morphology of which agree well with recent simulations and experiments.
	
We begin with the dynamics of a passive, impermeable membrane interacting with a suspension, active or passive~\cite{Sahu2017, Sahu2020}. 
The membrane is immersed in a $d$-dimensional solvent and is flat at rest (a line in $d = 2$ or a plane in $d=3$ with no spontaneous curvature).  
The normal coordinate to the flat membrane is $z$, and $\mathbf{x}$ denotes the $(d-1)$ coordinates along it.
Deviations from flat are described by the height function $h(\mathbf{x}, t)$~\cite{Kamien2002}.

The suspension occupies $z \geq h(\mathbf{x}, t)$ and exerts a pressure on the membrane that depends on its internal mechanics and how the two are coupled.
Our goal is to understand how this interaction affects the membrane's dynamics in the overdamped (no inertia) and  linear (small deviation ${|\bm{\nabla}h|\ll 1}$) limit, governed by~\cite{Granek1997}:
\begin{equation}
\label{eq:EqnMot}
	\partial_t \tilde{h}(\mathbf{q},t) = \tilde{M}(\mathbf{q}) \left[\tilde{\xi}(\mathbf{q},t) - \delta \mathcal{H}/\delta \tilde{h}(\mathbf{q},t) -\tilde{P}(\mathbf{q},t)\right].
\end{equation}
Here, ${\tilde{h}(\textbf{q})= \int d\mathbf{x}e^{i \mathbf{q}\cdot \mathbf{x}} h(\mathbf{x})}$, denotes the Fourier transform (FT) in $\mathbf{x}$ and introduces the wave vector $\mathbf{q}$. 
The left-hand side is the FT of the membrane's local $z$-velocity. 
Energy dissipation as the membrane moves through the solvent gives rise to the mobility, $\tilde{M}(\mathbf{q})$, which is the FT of the $zz$-component of the Oseen tensor if hydrodynamic interactions between the suspension and membrane are ignored~\cite{Happel1973,Kim2005,Doi1986, Note1}.
The bracketed terms on the right-hand side are the non-dissipative, $\hat{\mathbf{z}}$-directed forces per unit $(d-1)$-dimensional area on the membrane:
The first, $\tilde{\xi}$, represents thermal fluctuations as a Gaussian, zero-mean, temporally white noise satisfying the fluctuation-dissipation theorem.
The second encodes the membrane's internal mechanics through the standard Helfrich Hamiltonian ${\mathcal{H} = \int d\mathbf{q} (\gamma q^2 + \kappa q^4) \tilde{h}^2/2}$~\cite{Helfrich1973, Canham1970}, where ${q = |\mathbf{q}|}$, $\gamma$ is the surface tension and $\kappa$ is the bending modulus. 
The final term is the pressure exerted by the suspension (in FT) which, if the particle-membrane interactions are short-ranged contact forces, is: ${P(\mathbf{x},t)= -\hat{\mathbf{n}} \cdot\bm{\sigma}(\mathbf{x},z=h(\mathbf{x},t))\cdot \hat{\mathbf{n}}}$~\cite{Doi1986}, where $\bm{\sigma}$ is the suspension stress and ${\hat{\mathbf{n}} = (\hat{\mathbf{z}}- \bm{\nabla}h)/(1+|\bm{\nabla}h|^2)^{1/2}}$ is the membrane normal~\cite{Kamien2002, Note2}.

Takatori and Sahu used Eq.~\eqref{eq:EqnMot} to study non-equilibrium fluctuations of vesicles loaded with bacteria~\cite{Takatori2020}.
However, instead of deriving $P$ from the stress tensor, they modeled it as an athermal stochastic variable caused by bacteria-membrane collisions. 
This captured the experimentally observed fluctuation spectrum scaling $\sim q^{-4}$ as $q \to 0$ (the equilibrium scaling is $q^{-2}$~\cite{Fisher2004}) but \textit{cannot} predict activity induced instabilities of otherwise stable membranes, because $P$ was assumed to be independent of $h$. 
From Eq.~\eqref{eq:EqnMot}, this implies $h$ always fluctuates around the stable minimum, $h = 0$. 
The pressure from a passive \textit{ideal} gas is independent of $h$, but this is generally not the case~\cite{Kamien2014, Dinsmore1996, Kaplan1994, Blokhuis2013, Urrutia2014}.
ABPs, ideal or interacting, are known to accumulate, and exert more pressure, on valleys in a boundary than its hills~\cite{Wensink2008, Elgeti2013, Lee2013, Ezhilan2015, Yan2015,Yan2018, Duzgun2018, Granek2024}.
Intuitively, this can lead to an instability for flexible membranes: the excess pressure grows the valleys, drawing in more active particles to grow them further.
This mechanism has qualitatively described instabilities in flexible filaments~\cite{Nikola2016, Granek2024}. Here, we develop a quantitative theory to understand activity induced membrane instabilities.

The key quantity is the suspension stress at the membrane surface. 
For a general overdamped suspension, this is found from the mechanical force balance: ${\bm{\nabla} \cdot \bm{\sigma} + \mathbf{b} = \bm{0}}$, where $\mathbf{b}$ are the body forces. 
We consider a system of ABPs acted on by:
a Brownian force resulting in translational diffusion with constant $D_T = k_B T/\zeta$, where $\zeta$ is the translational drag coefficient associated with the solvent drag force, a fixed propulsion force $\zeta v_0$, where $v_0$ is the ``swim speed'', in a direction $\hat{\mathbf{u}}$ that undergoes rotational diffusion with re-orientation time ${\tau_r = D_r^{-1}}$, and conservative, pairwise, repulsive inter-particle forces. 
Mechanical balance in this case gives~\cite{Omar2023}
\begin{equation}
	\label{eq:Jrho}
	\bm{\nabla}\cdot \bm{\sigma} + \mathbf{F} - \zeta \mathbf{J}_{\rho} = \bm{0},
\end{equation}
where $\mathbf{J}_{\rho}$ is the number density current. 
The non-conservative active and drag force densities appear as the body forces, ${\mathbf{F} = \zeta v_0 \mathbf{m}}$ and  $\zeta \mathbf{J}_{\rho}$ respectively, where $\mathbf{m}$ is the polarization (the local average of $\hat{\mathbf{u}}$).
The Brownian (ideal-gas) and conservative inter-particle forces contribute to $\bm{\sigma}$~\cite{Yan2015Swim,Epstein2019,Omar2020}.
The force exerted by the membrane enforces no-flux at its surface.

Generally, $\bm{\sigma}$ includes contributions from density gradients. 
However, when $\ell_0/b \gtrsim 1$ where ${\ell_0 = v_0 \tau_r}$ is the ABP ``run length'', these are negligible in comparison to similar terms appearing in $\textbf{F}$~\cite{Omar2023}, allowing $\bm{\sigma}$ to be approximated by the ``conservative pressure''~\cite{Note3}
: ${\bm{\sigma} \approx - P_{\rm c}(\rho) \mathbf{I}}$, where $\mathbf{I}$ is the identity. 
Here, $P_{\rm c}$ is specified by an ``equation of state'' that depends only on $\rho$, and may be obtained from simulations in the absence of a microscopic theory~\cite{Mallory2021}. 
In the non-interacting limit, ${P_{\rm c} \to k_B T \rho}$, while at the maximum packing density for hard spheres, $\rho_{\rm max}$, it diverges.
The pressure exerted by the ABPs on the membrane is ${P(\mathbf{x},t)=  P_{\rm c}(\mathbf{x},z=h(\mathbf{x},t),t)}$, and is solely determined by the local density of active particles in contact with membrane.
While the active body force, $\mathbf{F}$, does not directly contribute to this pressure, it impacts it through its influence on $\rho$, as we shall see.
On small length scales, $P$ should be treated as a stochastic variable, as suggested in Ref.~\cite{Takatori2020}.
However, we focus on large-scale deformations driven by \textit{many} ABPs that are better handled using a coarse-grained, deterministic description. 
While this prevents us from accurately describing the membrane's fluctuation spectrum, we can determine its stability, which is controlled by the deterministic terms in Eq.~\eqref{eq:EqnMot}.

The mechanical balance Eq.~\eqref{eq:Jrho} makes it clear that to determine the surface density we must understand \textit{both} the complete density field, $\rho$, \textit{and} the polarization $\mathbf{m}$. 
Each is governed by a continuity equation:
\begin{equation}
    \label{eq:ctyrho}
		\partial_t \rho = - \bm{\nabla} \cdot \mathbf{J}_{\rho}, \ \ \text{and} \ \ \partial_t \mathbf{m} = - (d-1) D_r \mathbf{m} - \bm{\nabla} \cdot \mathbf{J}_{m}.
\end{equation}
The sink $\propto \mathbf{m}$ in the second represents relaxation due to rotational diffusion, the constitutive equation for $\mathbf{J}_{\rho}$ is given by Eq.~\eqref{eq:Jrho} and $\mathbf{J}_{m}$ is the polarization current, for which we use the following phenomenological equation that may be motivated from microscopic principles~\cite{Omar2023}:
\begin{equation}
	\label{eq:Jm}
	\mathbf{J}_{m} =- D_T \bm{\nabla} \mathbf{m} + v_0 \rho \mathcal{V}(\rho) \mathbf{I}/d.
\end{equation}
This neglects the nematic order of the ABPs, a good approximation outside \textit{rigid} walls with long-wavelength variations ($q \lesssim b^{-1}$) \cite{Yan2015,Yan2018}, and supposes inter-particle interactions can be accounted for through $\mathcal{V}(\rho)$, representing the dimensionless local average ABP swim speed. 
At low densities or in the absence of interactions, they swim freely: ${\mathcal{V} \to 1}$. 
At high densities they are trapped: $\mathcal{V}(\rho_{\rm max}) \to 0$. 
The second term in Eq.~\eqref{eq:Jm} leads to an ``active pressure'', ${P_{\rm act}(\rho) = \zeta \ell_0 v_0 \rho \mathcal{V}(\rho)/[d(d-1)] =k_B T\alpha^2 \rho \mathcal{V}(\rho)}$, due to forces generated by the particles' swimming, whose scale is set by both their density and activity~\cite{Takatori2014, Omar2020}, defined here by: ${\alpha= \ell_0/\Lambda_d [d (d-1)^2]^{-1/2}}$, where ${\Lambda_d = [D_T \tau_r /(d-1)]^{1/2}}$ is the ``diffusion length''.

It is well known that above certain activities and densities, homogeneous interacting ABP suspensions undergo ``Motility Induced Phase Separation'' (MIPS) into regions of high and low density~\cite{Cates2015}. 
ABPs also accumulate at walls, potentially rendering a suspension stable at bulk density $\rho_{\infty}$ unstable near the boundary.  
We assess the onset of MIPS using the mechanical \textit{spinodal} criterion ${\dot{P}_{\rm c}(\rho) + \dot{P}_{\rm act}(\rho) \leq 0}$ (dots denote $\rho$ derivatives), identifying the linear instability of the homogeneous suspension~\cite{Omar2023}.
Understanding the response of a membrane interacting with phase separating ABPs, including nucleation that can occur \textit{between} the spinodal and binodal, is a rich but complicated problem. 
We focus on suspensions that are linearly stable everywhere, including at the wall-enhanced density~\cite{Note4}
and neglect nucleation, while still finding interesting  behavior. 

Equations~(\ref{eq:EqnMot}-\ref{eq:Jm}) are closed, involving, $h$, $\rho$ and $\mathbf{m}$ \textit{only}, but are non-linear and difficult to solve analytically.  
To proceed, we make an additional approximation based on the separation of timescales between ABP relaxation and that of low-tension membranes. 
From Eqs.~\eqref{eq:EqnMot} \&~\eqref{eq:ctyrho}, a membrane perturbation with $q \ll b^{-1}$ relaxes at a rate $\propto q^3$ when $\gamma \approx 0$, while it relaxes more rapidly in the ABP density, $\propto q^2$.
This suggests an ``adiabatic approximation'': at each instant, the ABPs are assumed to be in steady state with the membrane profile. 
A finite surface tension requires $\alpha$ to be sufficiently large for this to hold, and it will describe wavenumbers in the range $\alpha^{-2} \lesssim b q \lesssim \alpha^2$ (see App.~\ref{app:Times} for detailed derivation).
For typical membrane properties~\cite{Vutukuri2020}, $\alpha \gtrsim 10$ is required for the approximation to be quantitatively accurate, but qualitative differences are not expected for smaller $\alpha$.

The adiabatic approximation removes the time derivatives from Eqs.~\eqref{eq:ctyrho}, leaving
\begin{equation}
	\label{eq:sseqns}
	\nabla^2 P_{\rm c} = \bm{\nabla} \cdot \mathbf{F}, \ \ \ \text{and}  \ \ \ \Lambda_d^{2}\nabla^2 \mathbf{F} - \mathbf{F} = \bm{\nabla} P_{\rm act}.
\end{equation}
These are closed equations for the density, via $P_{\rm c}$ and $P_{\rm act}$, and polarization, via $\mathbf{F}$, subject to the following boundary conditions (BCs): 
an apolar, homogeneous bulk, ${\rho(z \to \infty) \to \rho_{\infty}}$ and ${\mathbf{m}(z \to \infty) \to \bm{0}}$, and zero normal flux at the impermeable membrane
\begin{equation}
	\label{eq:BCs}	\hat{\mathbf{n}} \cdot \left(\bm{\nabla}P_{\rm c} - \mathbf{F} \right)\Big\lvert_{z=h}=0 \ \ \text{and} \ \ \hat{\mathbf{n}} \cdot \left(\Lambda_d^2 \bm{\nabla}\mathbf{F} -  P_{\rm act} \right)\Big\lvert_{z=h} = \bm{0}.\end{equation} 
Consistent with Eq.~\eqref{eq:EqnMot}, we aim to find $P_c(z=h)$ to first order in $|\bm{\nabla} h|$ from Eqs.~\eqref{eq:sseqns} \&~\eqref{eq:BCs}. 
To that end, we introduce ${\xi(\mathbf{x},z) = z-h(\mathbf{x},t)}$, measuring position above the \textit{deformed} membrane, and the associated coordinates along it ${\bm{\mu}(\mathbf{x},z) = \mathbf{x}}$~\cite{Note5}.
These allow BCs to be imposed at $\xi = 0$, at a cost of derivatives with respect to $(\mathbf{x},z)$ becoming more complicated: in the small deviation limit, we need only keep the resulting terms linear in $|\bm{\nabla} h|$ [App.~\ref{app:solcon} Eq.~\eqref{eq:appgeo}]. 
We expand $P_{\rm c}$ and $\mathbf{F}$ to the same order as: 
    \begin{subequations}
	\label{eq:appexp}
	\begin{equation}
 	\label{eq:psiexp}
		P_{\rm c}(\mathbf{x},z) \approx P_{\rm c}^{(0)}(\xi) + h(\bm{\mu},t) \partial_{\xi}P_c^{(0)} + p_{\rm c}(\bm{\mu},\xi),
	\end{equation}
	\begin{equation}
	\label{eq:Piexp}
		\mathbf{F}(\mathbf{x},z) \approx F^{(0)}(\xi) \hat{\mathbf{z}} + h(\bm{\mu},t) \partial_{\xi}F^{(0)} \hat{\mathbf{z}} + \bm{f}(\bm{\mu},\xi),
	\end{equation}
\end{subequations}
where $P_{\rm c}^{(0)}(z)$ and ${\mathbf{F}^{(0)} = F^{(0)}(z)\hat{\mathbf{z}}}$ are the solutions to Eq.~\eqref{eq:sseqns} for ${h=0}$, well established without interactions~\cite{Yan2015,Duzgun2018} and found as power series near $\xi = 0$ otherwise [Eq.~\eqref{eq:appflat}].
By symmetry, these depend only on $z$, and $\mathbf{F}^{(0)}$ only has a $z$-component. 
The first terms distort the flat-wall solutions to match the membrane profile, while $p_{\rm c}$ and $\bm{f}$ are the first-order corrections needed for Eqs.~\eqref{eq:appexp} to satisfy Eqs.~\eqref{eq:sseqns}.
Both $P_{\rm act}$ and $P_{\rm c}$ are functions of $\rho$, so their first-order corrections are related by the chain rule:
\begin{equation}
\label{eq:chi}
p_{\rm act}(\bm{\mu},\xi) = \frac{\dot{P}{\rm act}(\rho_0(\xi))}{\dot{P}{\rm c}(\rho_0(\xi))} p_{\rm c}(\bm{\mu},\xi) \equiv \chi(\xi) p_{\rm c}(\bm{\mu},\xi).
\end{equation}
Here $\rho_0$ is the flat-wall density profile, which determines $\chi(\xi)$.
In the non-interacting limit $\chi = \alpha^2$, allowing exact solutions [Eqs.~\eqref{eq:appnonint}].
With interactions, $\chi$ is a function of position, complicating our analysis. 

Equations~(\ref{eq:sseqns}-\ref{eq:chi}) lead to closed, linear equations for $p_{\rm c}$ and $\bm{f}$ found, along with their solutions, in App.~\ref{app:solcon}. 
The solution has three independent functions (one for $p_{\rm c}$ and two for the components of $\bm{f}$ parallel and perpendicular to the membrane) which, after FT in $\bm{x}$, appear with coefficients $c_1(\mathbf{q})$, $c_2(\mathbf{q})$ and $c_3(\mathbf{q})$, defined such that: ${\widetilde{p}_c(\mathbf{q},\xi=0) = c_1(\mathbf{q})}$.
The BCs in Eq.~\eqref{eq:BCs} give linear algebraic equations for $c_{1 \to 3}$. 

The pressure at the membrane surface is determined by the function, $\psi(\xi)$, multiplying $c_1$. 
This controls its decay of $P_{c}$ into the bulk and solves [Eqs.~\eqref{eq:pcapp} \&~\eqref{eq:psiphi}]
\begin{equation}
	\label{eq:feqn}
	\Lambda_d^2 \partial_{\xi}^2\psi- \left(1+\Lambda_d^2 q^2 +\chi(\xi)\right) \psi= 0,
\end{equation}
with $\psi(\xi=0)=1$ and $\psi(\xi \to \infty)=0$. 
This can be solved exactly without, but not with, interactions (App.~\ref{app:solcon}).
However, note that Eq.~\eqref{eq:feqn} is a Schr\"odinger equation for a particle of energy $-(1+\Lambda_d^2 q^2)$ in a potential $\chi(\xi)$.   
No MIPS and passive stability means ${\chi(\xi) > -(1+\Lambda_d^2 q^2)}$: in quantum mechanical language, the particle is in a ``classically forbidden'' region and the ``wavefunction'', $\psi$, exponentially decays. 

This analogy to quantum mechanics is useful. When $\chi$ (and hence $\rho_0$) is slowly varying, $\psi$ is accurately captured by the well-known Wentzel-Kramers-Brillouin (WKB) approximation~\cite{Bender1978, Dingle1973}.
The difference between $\rho_0(0)$ and $\rho_{\infty}$ is controlled by $P_{\rm act}$ in bulk ${\propto \alpha^2 \rho_{\infty} \mathcal{V}(\rho_{\infty})}$.
This can be large, even for moderate $\alpha$, at low $\rho_{\infty}$, resulting in steep concentration gradients and a poor WKB approximation.
A better approach here is to perturb around the non-interacting solution.
At higher $\rho_{\infty}$, interactions reduce the bulk active pressure and flatten $\rho_0(\xi)$, improving the WKB approximation. 
Thus, both high and low densities are tractable. 

The resulting pressure from the ABPs is, 
\begin{equation}
\label{eq:pext}
    \begin{split}
	\widetilde{P}(\mathbf{q},t) \approx P_{\rm c}(\rho_{0}(0)) \delta(\mathbf{q}) +\sum_{n=1}^{4} q^n X_n \tilde{h}(\mathbf{q},t),
    \end{split}
\end{equation}
where the coefficients $X_n$ can be found explicitly as functions of $\alpha$ and $\rho_{\infty}$ (see Supplemental Material~\cite{SeeDetails}).
We retain terms up to $q^4$, consistent with Eq.~\eqref{eq:EqnMot}.
The first term only affects the $q=0$ mode and is independent of $\tilde{h}$. 
This represents the uniform pressure exerted by ABPs on the membrane, equivalent to osmotic pressure, now modified by activity. 
The remaining linear in $\tilde{h}$ terms can be absorbed into an effective Hamiltonian in Eq.~\eqref{eq:EqnMot}, renormalizing the apparent membrane mechanics. 
Specifically, $X_2$ and $X_4$ renormalize the surface tension and bending modulus respectively:
\begin{equation}
\label{eq:effprops}
		\gamma_{\rm eff} = \gamma +  X_2(\alpha,\rho_{\infty}), \quad \text{and} \quad \kappa_{\rm eff} = \kappa + X_4(\alpha,\rho_{\infty}),
	\end{equation}    
showing that when $X_2$ and/or $X_4$ are negative, ABPs destabilize the membrane by decreasing its effective surface tension and/or bending modulus. 
$X_1$ and $X_3$ encode new non-local interactions from the redistribution of ABPs along the membrane~\cite{SeeDetails}, absent in the standard Helfrich Hamiltonian and when the membrane interacts with an equilibrium fluid with short-range interactions~\cite{Blokhuis2013,Urrutia2014}. 
Without interactions, $X_1$ \textit{vanishes}~\cite{Note6}.

The membrane's stability is probed by substituting Eq.~\eqref{eq:pext} into Eq.~\eqref{eq:EqnMot}. 
A mode of wavenumber $q$ is unstable if the resulting deterministic terms on the right-hand side are positive.
The instability morphology depends on which $q$ are unstable and grow most rapidly with four qualitatively distinct possibilities: 
(I) a finite wavelength is most unstable with long wavelengths ($q \to 0$) also unstable;
(II) same as (I) but ${q \to 0}$ are \textit{stable};
(III) short wavelengths ($q \to \infty$) are the most unstable, dominating other instabilities;
(IV) both long and short wavelengths unstable, but possibly a small finite range of stable wavelengths~\cite{Note7}.
The new $X_1$ and $X_3$ terms enable type II instabilities where a finite wavelength grows over all others, akin to Turing patterns~\cite{Maini2019} or modulated lipid domains and curvature in bilayer membranes~\cite{Yu2025}.

\begin{figure}\includegraphics[width=8.8cm]{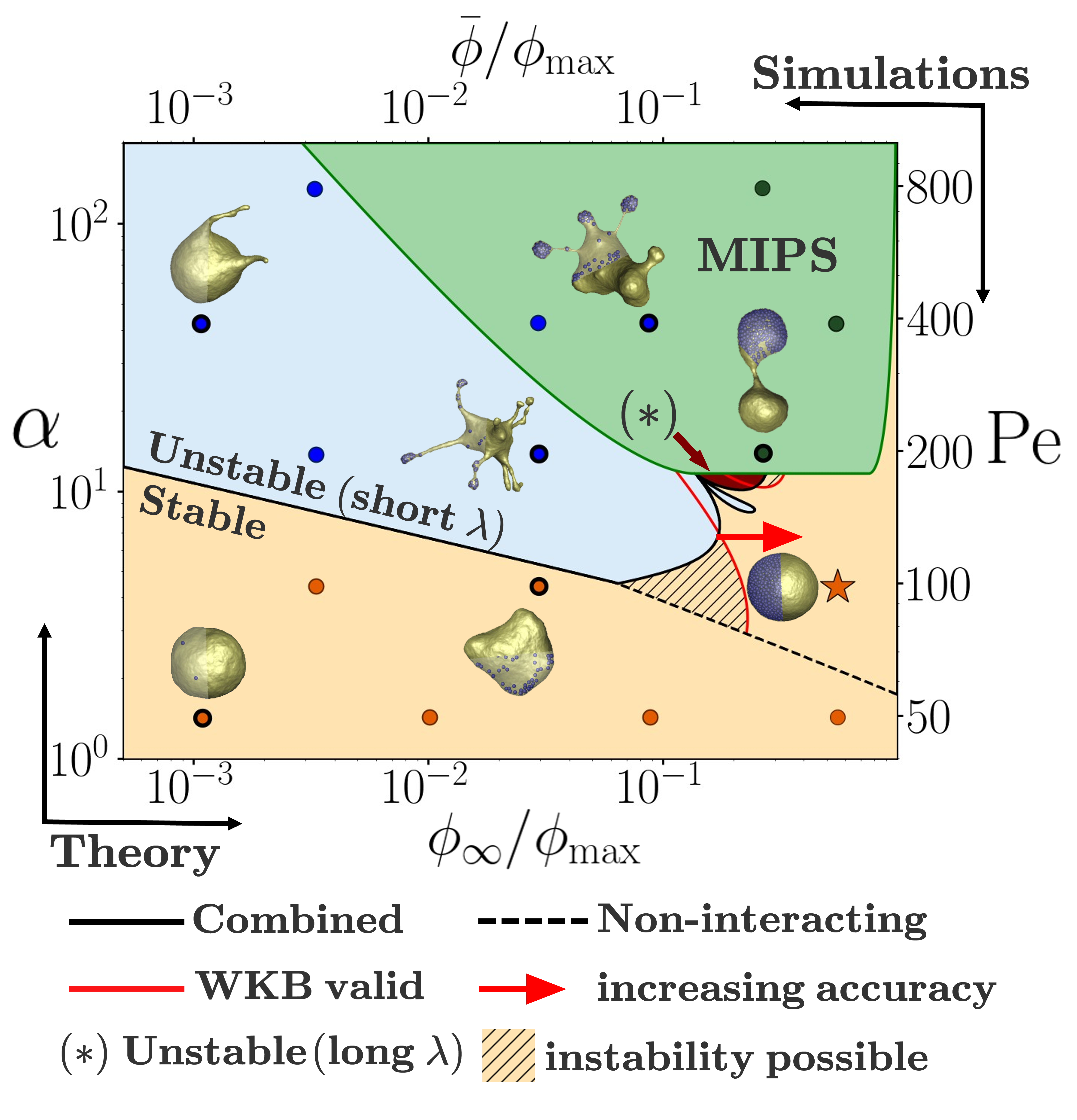}
\caption{\label{fig:Fig}\textit{Theory (left axes)}: Stability as a function of $\phi_\infty/\phi_{\rm max}$ and $\alpha$ in $d=3$. 
Colors denote stable (orange), short-wavelength ($\lambda$) unstable (blue), long $\lambda$ unstable (dark red), and MIPS (green) regimes. 
The solid black line is the non-interacting stability boundary at low $\phi_\infty$, replaced by the interacting (WKB) prediction, accurate right of the red curve, when they cross at higher $\phi_{\infty}$ where the non-interacting line continues dashed. 
In the hashed region, WKB is unreliable.
\textit{Simulations (right axes)}: Results from Ref.~\cite{Vutukuri2020} shown versus $\overline{\phi}/\phi_{\rm max}$ and ${\rm Pe}$. 
Points (outlined for snapshots) colored as in~\cite{Vutukuri2020}: orange (``fluctuating''), blue (``tethering''), and green (``bola''). 
Two sets of axes allow qualitative comparison between different membrane geometries.}
\end{figure}

Figure~\ref{fig:Fig} maps the predicted membrane instabilities as a function of ABP density, expressed as the reduced bulk volume fraction, $\phi_\infty/\phi_{\rm max}$,  and activity in $d=3$.
The equations of state, $P_c(\rho)$ and $P_{\rm act}(\rho)$, and $\phi_{\rm max} = 0.645$ are taken from hard-sphere ABP simulations~\cite{Omar2021, Omar2023}.
To compare with the dry ABP-laden spherical vesicle simulations of Vutukuri et al.~\cite{Vutukuri2020}, we match their membrane parameters. 
They considered relatively small (radius of $8b$) vesicles whose closed geometry means that ABP conservation is important: bulk density is ill-defined so they report overall ABP volume fraction, $\overline{\phi}$.
Since our theory for an infinite membrane doesn't quantitatively apply to small vesicles, we plot the simulation data on their own axes, aligned with the theory such that the onset of instability occurs at approximately the same location.
This allows a \textit{qualitative} comparison of the two stability maps.
In simulations, ABP activity is measured by the Péclet number $\mathrm{Pe}=v_0 b/D_T$.

The features of the theoretical and simulation stability maps agree excellently. 
In particular, at low volume fractions and activities, both show the membrane is stable and that increasing the activity here produces short wavelength instabilities (type III): the shortest wavelengths described by the theory are unstable, and simulations show long tendrils a few particle radii wide. 

The boundary between the stable and unstable regions is shown in solid black and is a combination of the non-interacting and interacting WKB solutions. 
For low $\phi_{\infty}/\phi_{\rm max}$, the non-interacting solution is expected to be accurate and provides the boundary, while the WKB approximation is poor.
At higher $\phi_{\infty}$, interactions are essential and WKB determines the stability boundary; the non-interacting solution continues as a dashed line.
The WKB approximation is valid to the right of the red line; the hashed region is where the WKB solution predicts stability, but its accuracy is uncertain.
Crucially, the interacting solution correctly captures the stability of vesicles at high $\phi_{\infty}$ seen in simulations (orange star), where the non-interacting solution incorrectly predicts instability.
Membrane stability at high $\phi_{\infty}$ is intuitive: when ABPs are trapped ($\mathcal{V}(\rho_{\rm max}) = 0$), they cannot destabilize the membrane. 

At high activities and volume fractions, MIPS occurs, either in bulk or due to accumulation at the membrane wall~\cite{SeeDetails}. 
While our theory is not applicable here, simulation results are consistent with phase separation driving membrane instability. 
Specifically, snapshots show high-density ABP clusters pushing droplets out of the lower-density vesicle or dividing it completely.
Near the MIPS boundary, WKB theory predicts a small region of long-wavelength instability.
This may be an artefact of the approximation, as it appears at the limit of its validity and is insensitive to membrane parameters~\cite{SeeDetails}, although it could also reflect genuine pre-transitional growth of long-wavelength fluctuations preceding phase separation.
\begin{figure}\includegraphics[width=8.5cm]{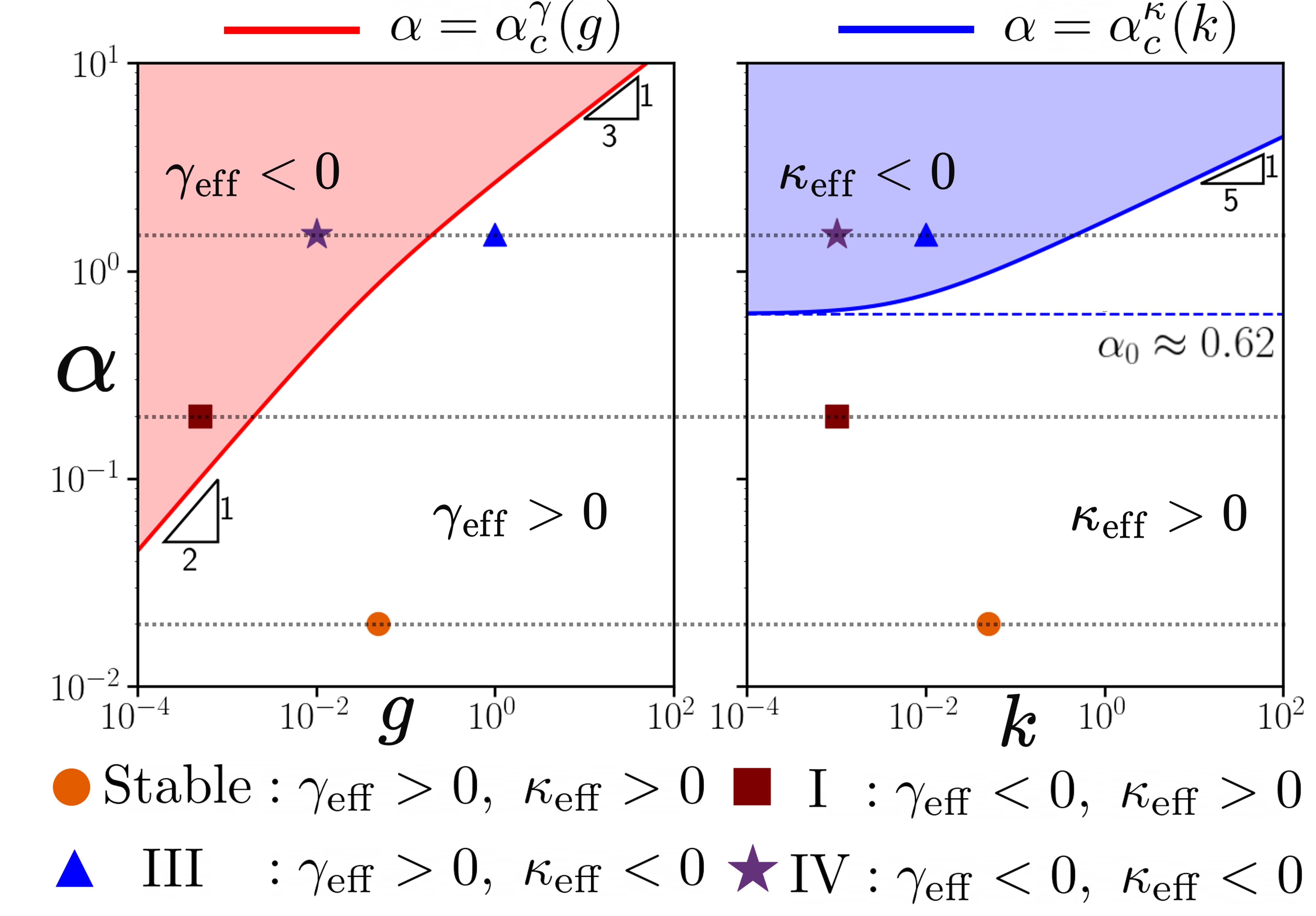}
	\caption{\label{fig:nonint} \textit{Left:} Critical activity $\alpha_c^{\gamma}$ above which $\gamma_{\rm eff}<0$ plotted against $g=\gamma V_d(b)/(\Lambda_d k_B T)$. 
    \textit{Right:} $\alpha_c^{\kappa}$ above which $\kappa_{\rm eff}<0$ vs. $k=\kappa V_d(b)/(\Lambda_d^3 k_B T)$.
    Both share a common vertical axis and are computed in the non-interacting limit at $\phi_\infty/\phi_{\rm max}=0.05$ with scalings at large and small $g$ and $k$ indicated.
    Tracing a horizontal line at fixed $\alpha$ across both panels immediately reveals the signs of $\gamma_{\rm eff}$ and $\kappa_{\rm eff}$ and the symbols illustrate the different instability types that arise.}
\end{figure}

The features of the stability diagram in Fig.~\ref{fig:Fig} depend on the bare membrane parameters, as expressed in Eqs.~\eqref{eq:effprops}. 
The Supplemental Material shows diagrams for a range of $\gamma$ and $\kappa$ but the main trends are captured by the non-interacting limit~\cite{SeeDetails}. 
There, $X_1$ vanishes, so the instability type is set by the signs of $\gamma_{\rm eff}$ and $\kappa_{\rm eff}$: (I) only $\gamma_{\rm eff}<0$, (III) only $\kappa_{\rm eff}<0$, and (IV) both negative.

Let $\alpha_c^{\gamma}$ and $\alpha_c^{\kappa}$ satisfy ${\gamma_{\rm eff}(\alpha_c^{\gamma}) =0}$, and ${\kappa_{\rm eff}(\alpha_c^{\kappa}) =0}$; above these activities, the membrane is unstable.  
In the non-interacting limit ($\chi = \alpha^2$), these can be determined analytically from the solutions to Eqs.~\eqref{eq:sseqns} \&~\eqref{eq:BCs} at fixed $\rho_{\infty}$.
From Eqs.~\eqref{eq:effprops} $\alpha_c^{\gamma}$ depends only on $\gamma$, while $\alpha_{c}^{\kappa}$ depends only on $\kappa$.
In Fig.~\ref{fig:nonint} we plot $\alpha_c^{\gamma}$ (red, left) and $\alpha_c^{\kappa}$ (blue, right) at $\phi_{\infty}/\phi_{\rm max} = 0.05$ as a function of the dimensionless membrane tension $g = \gamma V_d(b)/(k_B T \Lambda_d)$ and bending modulus $k = \kappa V_d(b)/(k_B T \Lambda_d^3)$ respectively, where $V_d(b)$ is $d$-dimensional spherical ABP volume. 
This shows membrane tension and/or bending modulus must be sufficiently small to see activity induced instabilities, consistent with experiments reporting instabilities only when membrane tension is reduced by orders of magnitude~\cite{Vutukuri2020, Takatori2020, Kim2025}.

For membrane parameters used in Fig.~\ref{fig:Fig} ($k \approx 154$, $g \approx 6$), only type III instabilities occur at low ABP densities. 
Figure~\ref{fig:nonint} shows how these instabilities can be tuned with the bare membrane properties: reducing $\gamma$ at fixed $\kappa$ allows 
long-wavelength (type I) instabilities to appear. 
We can also extract useful scaling relations (shown in Fig.~\ref{fig:nonint}): $\alpha_c^{\gamma} \sim (g/\phi_{\infty})^{1/3}$ for $g \gg 1$ and $\alpha_c^{\kappa} \sim (k/\phi_{\infty})^{1/5}$ for $k \gg 1$, while in the opposite limits $\alpha_c^{\gamma} \sim (g/\phi_{\infty})^{1/2}$ and $\alpha_c^{\kappa} \sim \alpha_0 \approx 0.62$.
The $\phi_{\infty}$ dependence follows from the linearity of $X_n$ in $\rho_{\infty}$ in the non-interacting limit.  

Interactions mean $X_1 \neq 0$, complicating how stability depends on membrane properties. 
In App.~\ref{app:MemProp} we quantify this by plotting where the interacting stability boundary departs from the non-interacting one as a function of bare membrane parameters.
The trends are: increasing $\gamma$ shifts the short-wavelength–unstable region to higher $\alpha$ and $\phi_{\infty}$, while increasing $\kappa$ shifts it to higher $\alpha$ but \textit{lower} $\phi_{\infty}$.

We have developed a theory for a passive membrane interacting with ABPs, yielding a stability diagram that captures the behavior seen in simulations and experiments across ABP density and activity. 
This shows how active stresses renormalize the membrane's surface tension and bending modulus, and generate long-range couplings along it.
Changes in membrane tension regulate important cellular processes: for example, a reduction in membrane tension in mouse embryonic stem cell development promotes endocytosis and drives early differentiation~\cite{Bergert2021,DeBelly2021}.
Our work shows how membrane mechanics are altered by its interactions with active suspensions, not just by the membrane's physical properties (composition, chemistry etc). 
This provides a direct physical route by which intra- or extracellular activity can modulate membrane tension, offering insight into the mechanics of these tension-dependent processes.

Our framework opens a route toward a broader theory of active-matter-driven membranes. 
A natural next step is to incorporate hydrodynamic interactions between particles, neglected here and in the simulations we compare against~\cite{Vutukuri2020}, but recently shown to strongly suppress MIPS~\cite{Zhou2026}. 
More generally, moving beyond the adiabatic and deterministic limits, particularly within the MIPS regime, may enable a unified multi-scale description of active suspensions coupled to deformable interfaces.

\textit{Acknowledgments - }  This work is supported by the U.S. Department of Energy, Office of Science, Office of Basic Energy Sciences, Materials Sciences and Engineering Division under Contract No. DE-AC02-05-CH11231 within the Adaptive Interfacial Assemblies Towards Structuring Liquids program (KCTR16). 


\begin{thebibliography}{65}%
\makeatletter
\providecommand \@ifxundefined [1]{%
 \@ifx{#1\undefined}
}%
\providecommand \@ifnum [1]{%
 \ifnum #1\expandafter \@firstoftwo
 \else \expandafter \@secondoftwo
 \fi
}%
\providecommand \@ifx [1]{%
 \ifx #1\expandafter \@firstoftwo
 \else \expandafter \@secondoftwo
 \fi
}%
\providecommand \natexlab [1]{#1}%
\providecommand \enquote  [1]{``#1''}%
\providecommand \bibnamefont  [1]{#1}%
\providecommand \bibfnamefont [1]{#1}%
\providecommand \citenamefont [1]{#1}%
\providecommand \href@noop [0]{\@secondoftwo}%
\providecommand \href [0]{\begingroup \@sanitize@url \@href}%
\providecommand \@href[1]{\@@startlink{#1}\@@href}%
\providecommand \@@href[1]{\endgroup#1\@@endlink}%
\providecommand \@sanitize@url [0]{\catcode `\\12\catcode `\$12\catcode `\&12\catcode `\#12\catcode `\^12\catcode `\_12\catcode `\%12\relax}%
\providecommand \@@startlink[1]{}%
\providecommand \@@endlink[0]{}%
\providecommand \url  [0]{\begingroup\@sanitize@url \@url }%
\providecommand \@url [1]{\endgroup\@href {#1}{\urlprefix }}%
\providecommand \urlprefix  [0]{URL }%
\providecommand \Eprint [0]{\href }%
\providecommand \doibase [0]{https://doi.org/}%
\providecommand \selectlanguage [0]{\@gobble}%
\providecommand \bibinfo  [0]{\@secondoftwo}%
\providecommand \bibfield  [0]{\@secondoftwo}%
\providecommand \translation [1]{[#1]}%
\providecommand \BibitemOpen [0]{}%
\providecommand \bibitemStop [0]{}%
\providecommand \bibitemNoStop [0]{.\EOS\space}%
\providecommand \EOS [0]{\spacefactor3000\relax}%
\providecommand \BibitemShut  [1]{\csname bibitem#1\endcsname}%
\let\auto@bib@innerbib\@empty
\bibitem [{\citenamefont {May}\ and\ \citenamefont {Machesky}(2001)}]{May2001}%
  \BibitemOpen
  \bibfield  {author} {\bibinfo {author} {\bibfnamefont {R.~C.}\ \bibnamefont {May}}\ and\ \bibinfo {author} {\bibfnamefont {L.~M.}\ \bibnamefont {Machesky}},\ }\bibfield  {title} {\bibinfo {title} {Phagocytosis and the actin cytoskeleton},\ }\href {https://doi.org/10.1242/jcs.114.6.1061} {\bibfield  {journal} {\bibinfo  {journal} {Journal of Cell Science}\ }\textbf {\bibinfo {volume} {114}},\ \bibinfo {pages} {1061} (\bibinfo {year} {2001})}\BibitemShut {NoStop}%
\bibitem [{\citenamefont {Pollard}\ and\ \citenamefont {Cooper}(2009)}]{Pollard2009}%
  \BibitemOpen
  \bibfield  {author} {\bibinfo {author} {\bibfnamefont {T.~D.}\ \bibnamefont {Pollard}}\ and\ \bibinfo {author} {\bibfnamefont {J.~A.}\ \bibnamefont {Cooper}},\ }\bibfield  {title} {\bibinfo {title} {Actin, a central player in cell shape and movement},\ }\href {https://doi.org/10.1126/science.1175862} {\bibfield  {journal} {\bibinfo  {journal} {Science}\ }\textbf {\bibinfo {volume} {326}},\ \bibinfo {pages} {1208} (\bibinfo {year} {2009})}\BibitemShut {NoStop}%
\bibitem [{\citenamefont {Rottner}\ and\ \citenamefont {Schaks}(2019)}]{Rottner2019}%
  \BibitemOpen
  \bibfield  {author} {\bibinfo {author} {\bibfnamefont {K.}~\bibnamefont {Rottner}}\ and\ \bibinfo {author} {\bibfnamefont {M.}~\bibnamefont {Schaks}},\ }\bibfield  {title} {\bibinfo {title} {Assembling actin filaments for protrusion},\ }\href {https://doi.org/10.1016/j.ceb.2018.09.004} {\bibfield  {journal} {\bibinfo  {journal} {Current Opinion in Cell Biology}\ }\textbf {\bibinfo {volume} {56}},\ \bibinfo {pages} {53} (\bibinfo {year} {2019})}\BibitemShut {NoStop}%
\bibitem [{\citenamefont {Welf}\ \emph {et~al.}(2020)\citenamefont {Welf}, \citenamefont {Miles}, \citenamefont {Huh}, \citenamefont {Sapoznik}, \citenamefont {Chi}, \citenamefont {Driscoll}, \citenamefont {Isogai}, \citenamefont {Noh}, \citenamefont {Weems}, \citenamefont {Pohlkamp}, \citenamefont {Dean}, \citenamefont {Fiolka}, \citenamefont {Mogilner},\ and\ \citenamefont {Danuser}}]{Welf2020}%
  \BibitemOpen
  \bibfield  {author} {\bibinfo {author} {\bibfnamefont {E.~S.}\ \bibnamefont {Welf}}, \bibinfo {author} {\bibfnamefont {C.~E.}\ \bibnamefont {Miles}}, \bibinfo {author} {\bibfnamefont {J.}~\bibnamefont {Huh}}, \bibinfo {author} {\bibfnamefont {E.}~\bibnamefont {Sapoznik}}, \bibinfo {author} {\bibfnamefont {J.}~\bibnamefont {Chi}}, \bibinfo {author} {\bibfnamefont {M.~K.}\ \bibnamefont {Driscoll}}, \bibinfo {author} {\bibfnamefont {T.}~\bibnamefont {Isogai}}, \bibinfo {author} {\bibfnamefont {J.}~\bibnamefont {Noh}}, \bibinfo {author} {\bibfnamefont {A.~D.}\ \bibnamefont {Weems}}, \bibinfo {author} {\bibfnamefont {T.}~\bibnamefont {Pohlkamp}}, \bibinfo {author} {\bibfnamefont {K.}~\bibnamefont {Dean}}, \bibinfo {author} {\bibfnamefont {R.}~\bibnamefont {Fiolka}}, \bibinfo {author} {\bibfnamefont {A.}~\bibnamefont {Mogilner}},\ and\ \bibinfo {author} {\bibfnamefont {G.}~\bibnamefont {Danuser}},\ }\bibfield  {title} {\bibinfo {title} {Actin-membrane release initiates cell protrusions},\ }\href
  {https://doi.org/10.1016/j.devcel.2020.11.024} {\bibfield  {journal} {\bibinfo  {journal} {Developmental Cell}\ }\textbf {\bibinfo {volume} {55}},\ \bibinfo {pages} {723} (\bibinfo {year} {2020})}\BibitemShut {NoStop}%
\bibitem [{\citenamefont {Pizarro-Cerdá}\ \emph {et~al.}(2016)\citenamefont {Pizarro-Cerdá}, \citenamefont {Charbit}, \citenamefont {Enninga}, \citenamefont {Lafont},\ and\ \citenamefont {Cossart}}]{Pizarro-Cerd2016}%
  \BibitemOpen
  \bibfield  {author} {\bibinfo {author} {\bibfnamefont {J.}~\bibnamefont {Pizarro-Cerdá}}, \bibinfo {author} {\bibfnamefont {A.}~\bibnamefont {Charbit}}, \bibinfo {author} {\bibfnamefont {J.}~\bibnamefont {Enninga}}, \bibinfo {author} {\bibfnamefont {F.}~\bibnamefont {Lafont}},\ and\ \bibinfo {author} {\bibfnamefont {P.}~\bibnamefont {Cossart}},\ }\bibfield  {title} {\bibinfo {title} {Manipulation of host membranes by the bacterial pathogens listeria, francisella, shigella and yersinia},\ }\href {https://doi.org/10.1016/j.semcdb.2016.07.019} {\bibfield  {journal} {\bibinfo  {journal} {Seminars in Cell \& Developmental Biology}\ }\textbf {\bibinfo {volume} {60}},\ \bibinfo {pages} {155} (\bibinfo {year} {2016})}\BibitemShut {NoStop}%
\bibitem [{\citenamefont {Friedrich}\ \emph {et~al.}(2012)\citenamefont {Friedrich}, \citenamefont {Hagedorn}, \citenamefont {Soldati-Favre},\ and\ \citenamefont {Soldati}}]{Friedrich2012}%
  \BibitemOpen
  \bibfield  {author} {\bibinfo {author} {\bibfnamefont {N.}~\bibnamefont {Friedrich}}, \bibinfo {author} {\bibfnamefont {M.}~\bibnamefont {Hagedorn}}, \bibinfo {author} {\bibfnamefont {D.}~\bibnamefont {Soldati-Favre}},\ and\ \bibinfo {author} {\bibfnamefont {T.}~\bibnamefont {Soldati}},\ }\bibfield  {title} {\bibinfo {title} {Prison break: Pathogens' strategies to egress from host cells},\ }\href {https://doi.org/10.1128/MMBR.00024-12} {\bibfield  {journal} {\bibinfo  {journal} {Microbiology and Molecular Biology Reviews}\ }\textbf {\bibinfo {volume} {76}},\ \bibinfo {pages} {707} (\bibinfo {year} {2012})}\BibitemShut {NoStop}%
\bibitem [{\citenamefont {Takatori}\ and\ \citenamefont {Sahu}(2020)}]{Takatori2020}%
  \BibitemOpen
  \bibfield  {author} {\bibinfo {author} {\bibfnamefont {S.~C.}\ \bibnamefont {Takatori}}\ and\ \bibinfo {author} {\bibfnamefont {A.}~\bibnamefont {Sahu}},\ }\bibfield  {title} {\bibinfo {title} {Active contact forces drive nonequilibrium fluctuations in membrane vesicles},\ }\href {https://doi.org/10.1103/PhysRevLett.124.158102} {\bibfield  {journal} {\bibinfo  {journal} {Physical Review Letters}\ }\textbf {\bibinfo {volume} {124}},\ \bibinfo {pages} {158102} (\bibinfo {year} {2020})}\BibitemShut {NoStop}%
\bibitem [{\citenamefont {Vutukuri}\ \emph {et~al.}(2020)\citenamefont {Vutukuri}, \citenamefont {Hoore}, \citenamefont {Abaurrea-Velasco}, \citenamefont {van Buren}, \citenamefont {Dutto}, \citenamefont {Auth}, \citenamefont {Fedosov}, \citenamefont {Gompper},\ and\ \citenamefont {Vermant}}]{Vutukuri2020}%
  \BibitemOpen
  \bibfield  {author} {\bibinfo {author} {\bibfnamefont {H.~R.}\ \bibnamefont {Vutukuri}}, \bibinfo {author} {\bibfnamefont {M.}~\bibnamefont {Hoore}}, \bibinfo {author} {\bibfnamefont {C.}~\bibnamefont {Abaurrea-Velasco}}, \bibinfo {author} {\bibfnamefont {L.}~\bibnamefont {van Buren}}, \bibinfo {author} {\bibfnamefont {A.}~\bibnamefont {Dutto}}, \bibinfo {author} {\bibfnamefont {T.}~\bibnamefont {Auth}}, \bibinfo {author} {\bibfnamefont {D.~A.}\ \bibnamefont {Fedosov}}, \bibinfo {author} {\bibfnamefont {G.}~\bibnamefont {Gompper}},\ and\ \bibinfo {author} {\bibfnamefont {J.}~\bibnamefont {Vermant}},\ }\bibfield  {title} {\bibinfo {title} {Active particles induce large shape deformations in giant lipid vesicles},\ }\href {https://doi.org/10.1038/s41586-020-2730-x} {\bibfield  {journal} {\bibinfo  {journal} {Nature}\ }\textbf {\bibinfo {volume} {586}},\ \bibinfo {pages} {52} (\bibinfo {year} {2020})}\BibitemShut {NoStop}%
\bibitem [{\citenamefont {Kim}\ \emph {et~al.}(2025)\citenamefont {Kim}, \citenamefont {Zhu}, \citenamefont {Forth}, \citenamefont {Xie}, \citenamefont {King}, \citenamefont {Helms}, \citenamefont {Ashby}, \citenamefont {Omar},\ and\ \citenamefont {Russell}}]{Kim2025}%
  \BibitemOpen
  \bibfield  {author} {\bibinfo {author} {\bibfnamefont {P.~Y.}\ \bibnamefont {Kim}}, \bibinfo {author} {\bibfnamefont {S.}~\bibnamefont {Zhu}}, \bibinfo {author} {\bibfnamefont {J.}~\bibnamefont {Forth}}, \bibinfo {author} {\bibfnamefont {G.}~\bibnamefont {Xie}}, \bibinfo {author} {\bibfnamefont {D.~A.}\ \bibnamefont {King}}, \bibinfo {author} {\bibfnamefont {B.~A.}\ \bibnamefont {Helms}}, \bibinfo {author} {\bibfnamefont {P.~D.}\ \bibnamefont {Ashby}}, \bibinfo {author} {\bibfnamefont {A.~K.}\ \bibnamefont {Omar}},\ and\ \bibinfo {author} {\bibfnamefont {T.~P.}\ \bibnamefont {Russell}},\ }\bibfield  {title} {\bibinfo {title} {Shape‐evolving structured liquids},\ }\href {https://advanced.onlinelibrary.wiley.com/doi/10.1002/adma.202500804} {\bibfield  {journal} {\bibinfo  {journal} {Advanced Materials}\ }\textbf {\bibinfo {volume} {37}} (\bibinfo {year} {2025})}\BibitemShut {NoStop}%
\bibitem [{\citenamefont {Sahu}\ \emph {et~al.}(2017)\citenamefont {Sahu}, \citenamefont {Sauer},\ and\ \citenamefont {Mandadapu}}]{Sahu2017}%
  \BibitemOpen
  \bibfield  {author} {\bibinfo {author} {\bibfnamefont {A.}~\bibnamefont {Sahu}}, \bibinfo {author} {\bibfnamefont {R.~A.}\ \bibnamefont {Sauer}},\ and\ \bibinfo {author} {\bibfnamefont {K.~K.}\ \bibnamefont {Mandadapu}},\ }\bibfield  {title} {\bibinfo {title} {Irreversible thermodynamics of curved lipid membranes},\ }\href {https://doi.org/10.1103/PhysRevE.96.042409} {\bibfield  {journal} {\bibinfo  {journal} {Physical Review E}\ }\textbf {\bibinfo {volume} {96}},\ \bibinfo {pages} {042409} (\bibinfo {year} {2017})}\BibitemShut {NoStop}%
\bibitem [{\citenamefont {Sahu}\ \emph {et~al.}(2020)\citenamefont {Sahu}, \citenamefont {Glisman}, \citenamefont {Tchoufag},\ and\ \citenamefont {Mandadapu}}]{Sahu2020}%
  \BibitemOpen
  \bibfield  {author} {\bibinfo {author} {\bibfnamefont {A.}~\bibnamefont {Sahu}}, \bibinfo {author} {\bibfnamefont {A.}~\bibnamefont {Glisman}}, \bibinfo {author} {\bibfnamefont {J.}~\bibnamefont {Tchoufag}},\ and\ \bibinfo {author} {\bibfnamefont {K.~K.}\ \bibnamefont {Mandadapu}},\ }\bibfield  {title} {\bibinfo {title} {Geometry and dynamics of lipid membranes: The scriven-love number},\ }\href {https://doi.org/10.1103/PhysRevE.101.052401} {\bibfield  {journal} {\bibinfo  {journal} {Physical Review E}\ }\textbf {\bibinfo {volume} {101}},\ \bibinfo {pages} {052401} (\bibinfo {year} {2020})}\BibitemShut {NoStop}%
\bibitem [{\citenamefont {Kamien}(2002)}]{Kamien2002}%
  \BibitemOpen
  \bibfield  {author} {\bibinfo {author} {\bibfnamefont {R.~D.}\ \bibnamefont {Kamien}},\ }\bibfield  {title} {\bibinfo {title} {The geometry of soft materials: a primer},\ }\href {https://doi.org/10.1103/RevModPhys.74.953} {\bibfield  {journal} {\bibinfo  {journal} {Reviews of Modern Physics}\ }\textbf {\bibinfo {volume} {74}},\ \bibinfo {pages} {953} (\bibinfo {year} {2002})}\BibitemShut {NoStop}%
\bibitem [{\citenamefont {Granek}(1997)}]{Granek1997}%
  \BibitemOpen
  \bibfield  {author} {\bibinfo {author} {\bibfnamefont {R.}~\bibnamefont {Granek}},\ }\bibfield  {title} {\bibinfo {title} {From semi-flexible polymers to membranes: Anomalous diffusion and reptation},\ }\href {https://doi.org/10.1051/jp2:1997214} {\bibfield  {journal} {\bibinfo  {journal} {Journal de Physique II}\ }\textbf {\bibinfo {volume} {7}},\ \bibinfo {pages} {1761} (\bibinfo {year} {1997})}\BibitemShut {NoStop}%
\bibitem [{\citenamefont {Happel}\ and\ \citenamefont {Brenner}(1983)}]{Happel1973}%
  \BibitemOpen
  \bibfield  {author} {\bibinfo {author} {\bibfnamefont {J.}~\bibnamefont {Happel}}\ and\ \bibinfo {author} {\bibfnamefont {H.}~\bibnamefont {Brenner}},\ }\href {https://doi.org/10.1007/978-94-009-8352-6} {\emph {\bibinfo {title} {Low Reynolds number hydrodynamics}}},\ Vol.~\bibinfo {volume} {1}\ (\bibinfo  {publisher} {Springer Netherlands},\ \bibinfo {year} {1983})\BibitemShut {NoStop}%
\bibitem [{\citenamefont {Kim}\ and\ \citenamefont {Karrila}(2005)}]{Kim2005}%
  \BibitemOpen
  \bibfield  {author} {\bibinfo {author} {\bibfnamefont {S.}~\bibnamefont {Kim}}\ and\ \bibinfo {author} {\bibfnamefont {S.~J.}\ \bibnamefont {Karrila}},\ }\href {https://books.google.co.uk/books?hl=en&lr=&id=aADEAgAAQBAJ&oi=fnd&pg=PP1&dq=kim+karrila+microhydrodynamics&ots=hH91igZ3Mr&sig=XVf21MGBQXz5H7d7iDv-LrkDgJg#v=onepage&q=kim karrila microhydrodynamics&f=false} {\emph {\bibinfo {title} {Microhydrodynamics : principles and selected applications}}}\ (\bibinfo  {publisher} {Dover Publications},\ \bibinfo {year} {2005})\BibitemShut {NoStop}%
\bibitem [{\citenamefont {Doi}\ and\ \citenamefont {Edwards}(1986)}]{Doi1986}%
  \BibitemOpen
  \bibfield  {author} {\bibinfo {author} {\bibfnamefont {M.}~\bibnamefont {Doi}}\ and\ \bibinfo {author} {\bibfnamefont {S.~F.}\ \bibnamefont {Edwards}},\ }\href {https://books.google.co.uk/books/about/The_Theory_of_Polymer_Dynamics.html?id=dMzGyWs3GKcC} {\emph {\bibinfo {title} {The theory of polymer dynamics}}}\ (\bibinfo  {publisher} {Oxford University Press},\ \bibinfo {year} {1986})\BibitemShut {NoStop}%
\bibitem [{Note1()}]{Note1}%
  \BibitemOpen
  \bibinfo {note} {Hydrodynamic interactions between the ABPs and the membrane can be approximately incorporated via an effective medium theory renormalizing $\protect \tilde {M}$ \cite {Edwards1974,Freed1974,Shaqfeh1990}.}\BibitemShut {Stop}%
\bibitem [{\citenamefont {Helfrich}(1973)}]{Helfrich1973}%
  \BibitemOpen
  \bibfield  {author} {\bibinfo {author} {\bibfnamefont {W.}~\bibnamefont {Helfrich}},\ }\bibfield  {title} {\bibinfo {title} {Elastic properties of lipid bilayers: Theory and possible experiments},\ }\href {https://doi.org/10.1515/znc-1973-11-1209} {\bibfield  {journal} {\bibinfo  {journal} {Zeitschrift für Naturforschung C}\ }\textbf {\bibinfo {volume} {28}},\ \bibinfo {pages} {693} (\bibinfo {year} {1973})}\BibitemShut {NoStop}%
\bibitem [{\citenamefont {Canham}(1970)}]{Canham1970}%
  \BibitemOpen
  \bibfield  {author} {\bibinfo {author} {\bibfnamefont {P.}~\bibnamefont {Canham}},\ }\bibfield  {title} {\bibinfo {title} {The minimum energy of bending as a possible explanation of the biconcave shape of the human red blood cell},\ }\href {https://doi.org/10.1016/S0022-5193(70)80032-7} {\bibfield  {journal} {\bibinfo  {journal} {Journal of Theoretical Biology}\ }\textbf {\bibinfo {volume} {26}},\ \bibinfo {pages} {61} (\bibinfo {year} {1970})}\BibitemShut {NoStop}%
\bibitem [{Note2()}]{Note2}%
  \BibitemOpen
  \bibinfo {note} {Generally, for finite-ranged interactions (e.g. electrostatic), $P$ must be determined by integrating over the entire suspension domain~\cite {Henderson1979}.}\BibitemShut {Stop}%
\bibitem [{\citenamefont {Fisher}(2004)}]{Fisher2004}%
  \BibitemOpen
  \bibfield  {author} {\bibinfo {author} {\bibfnamefont {M.~E.}\ \bibnamefont {Fisher}},\ }\bibinfo {title} {Interfaces: fluctuations, interactions and related transitions},\ in\ \href {https://doi.org/10.1142/9789812565518_0002} {\emph {\bibinfo {booktitle} {Statistical Mechanics of Membranes and Surfaces}}},\ \bibinfo {editor} {edited by\ \bibinfo {editor} {\bibfnamefont {D.}~\bibnamefont {Nelson}}, \bibinfo {editor} {\bibfnamefont {T.}~\bibnamefont {Piran}},\ and\ \bibinfo {editor} {\bibfnamefont {S.}~\bibnamefont {Weinberg}}}\ (\bibinfo  {publisher} {World Scientific},\ \bibinfo {year} {2004})\ pp.\ \bibinfo {pages} {19--47}\BibitemShut {NoStop}%
\bibitem [{\citenamefont {Kamien}(2007)}]{Kamien2014}%
  \BibitemOpen
  \bibfield  {author} {\bibinfo {author} {\bibfnamefont {R.~D.}\ \bibnamefont {Kamien}},\ }\bibinfo {title} {Entropic attraction and ordering},\ in\ \href {https://doi.org/10.1002/9783527682300.ch1} {\emph {\bibinfo {booktitle} {Soft Matter}}},\ \bibinfo {editor} {edited by\ \bibinfo {editor} {\bibfnamefont {G.}~\bibnamefont {Gompper}}\ and\ \bibinfo {editor} {\bibfnamefont {M.}~\bibnamefont {Schick}}}\ (\bibinfo  {publisher} {Wiley},\ \bibinfo {year} {2007})\ pp.\ \bibinfo {pages} {1--40}\BibitemShut {NoStop}%
\bibitem [{\citenamefont {Dinsmore}\ \emph {et~al.}(1996)\citenamefont {Dinsmore}, \citenamefont {Yodh},\ and\ \citenamefont {Pine}}]{Dinsmore1996}%
  \BibitemOpen
  \bibfield  {author} {\bibinfo {author} {\bibfnamefont {A.~D.}\ \bibnamefont {Dinsmore}}, \bibinfo {author} {\bibfnamefont {A.~G.}\ \bibnamefont {Yodh}},\ and\ \bibinfo {author} {\bibfnamefont {D.~J.}\ \bibnamefont {Pine}},\ }\bibfield  {title} {\bibinfo {title} {Entropic control of particle motion using passive surface microstructures},\ }\href {https://doi.org/10.1038/383239a0} {\bibfield  {journal} {\bibinfo  {journal} {Nature}\ }\textbf {\bibinfo {volume} {383}},\ \bibinfo {pages} {239} (\bibinfo {year} {1996})}\BibitemShut {NoStop}%
\bibitem [{\citenamefont {Kaplan}\ \emph {et~al.}(1994)\citenamefont {Kaplan}, \citenamefont {Rouke}, \citenamefont {Yodh},\ and\ \citenamefont {Pine}}]{Kaplan1994}%
  \BibitemOpen
  \bibfield  {author} {\bibinfo {author} {\bibfnamefont {P.~D.}\ \bibnamefont {Kaplan}}, \bibinfo {author} {\bibfnamefont {J.~L.}\ \bibnamefont {Rouke}}, \bibinfo {author} {\bibfnamefont {A.~G.}\ \bibnamefont {Yodh}},\ and\ \bibinfo {author} {\bibfnamefont {D.~J.}\ \bibnamefont {Pine}},\ }\bibfield  {title} {\bibinfo {title} {Entropically driven surface phase separation in binary colloidal mixtures},\ }\href {https://doi.org/10.1103/PhysRevLett.72.582} {\bibfield  {journal} {\bibinfo  {journal} {Physical Review Letters}\ }\textbf {\bibinfo {volume} {72}},\ \bibinfo {pages} {582} (\bibinfo {year} {1994})}\BibitemShut {NoStop}%
\bibitem [{\citenamefont {Blokhuis}(2013)}]{Blokhuis2013}%
  \BibitemOpen
  \bibfield  {author} {\bibinfo {author} {\bibfnamefont {E.~M.}\ \bibnamefont {Blokhuis}},\ }\bibfield  {title} {\bibinfo {title} {Existence of a bending rigidity for a hard-sphere liquid near a curved hard wall: Validity of the hadwiger theorem},\ }\href {https://doi.org/10.1103/PhysRevE.87.022401} {\bibfield  {journal} {\bibinfo  {journal} {Physical Review E}\ }\textbf {\bibinfo {volume} {87}},\ \bibinfo {pages} {022401} (\bibinfo {year} {2013})}\BibitemShut {NoStop}%
\bibitem [{\citenamefont {Urrutia}(2014)}]{Urrutia2014}%
  \BibitemOpen
  \bibfield  {author} {\bibinfo {author} {\bibfnamefont {I.}~\bibnamefont {Urrutia}},\ }\bibfield  {title} {\bibinfo {title} {Bending rigidity and higher-order curvature terms for the hard-sphere fluid near a curved wall},\ }\href {https://doi.org/10.1103/PhysRevE.89.032122} {\bibfield  {journal} {\bibinfo  {journal} {Physical Review E}\ }\textbf {\bibinfo {volume} {89}},\ \bibinfo {pages} {032122} (\bibinfo {year} {2014})}\BibitemShut {NoStop}%
\bibitem [{\citenamefont {Wensink}\ and\ \citenamefont {Löwen}(2008)}]{Wensink2008}%
  \BibitemOpen
  \bibfield  {author} {\bibinfo {author} {\bibfnamefont {H.~H.}\ \bibnamefont {Wensink}}\ and\ \bibinfo {author} {\bibfnamefont {H.}~\bibnamefont {Löwen}},\ }\bibfield  {title} {\bibinfo {title} {Aggregation of self-propelled colloidal rods near confining walls},\ }\href {https://doi.org/10.1103/PhysRevE.78.031409} {\bibfield  {journal} {\bibinfo  {journal} {Physical Review E}\ }\textbf {\bibinfo {volume} {78}},\ \bibinfo {pages} {031409} (\bibinfo {year} {2008})}\BibitemShut {NoStop}%
\bibitem [{\citenamefont {Elgeti}\ and\ \citenamefont {Gompper}(2013)}]{Elgeti2013}%
  \BibitemOpen
  \bibfield  {author} {\bibinfo {author} {\bibfnamefont {J.}~\bibnamefont {Elgeti}}\ and\ \bibinfo {author} {\bibfnamefont {G.}~\bibnamefont {Gompper}},\ }\bibfield  {title} {\bibinfo {title} {Wall accumulation of self-propelled spheres},\ }\href {https://doi.org/10.1209/0295-5075/101/48003} {\bibfield  {journal} {\bibinfo  {journal} {EPL (Europhysics Letters)}\ }\textbf {\bibinfo {volume} {101}},\ \bibinfo {pages} {48003} (\bibinfo {year} {2013})}\BibitemShut {NoStop}%
\bibitem [{\citenamefont {Lee}(2013)}]{Lee2013}%
  \BibitemOpen
  \bibfield  {author} {\bibinfo {author} {\bibfnamefont {C.~F.}\ \bibnamefont {Lee}},\ }\bibfield  {title} {\bibinfo {title} {Active particles under confinement: aggregation at the wall and gradient formation inside a channel},\ }\href {https://doi.org/10.1088/1367-2630/15/5/055007} {\bibfield  {journal} {\bibinfo  {journal} {New Journal of Physics}\ }\textbf {\bibinfo {volume} {15}},\ \bibinfo {pages} {055007} (\bibinfo {year} {2013})}\BibitemShut {NoStop}%
\bibitem [{\citenamefont {Ezhilan}\ \emph {et~al.}(2015)\citenamefont {Ezhilan}, \citenamefont {Alonso-Matilla},\ and\ \citenamefont {Saintillan}}]{Ezhilan2015}%
  \BibitemOpen
  \bibfield  {author} {\bibinfo {author} {\bibfnamefont {B.}~\bibnamefont {Ezhilan}}, \bibinfo {author} {\bibfnamefont {R.}~\bibnamefont {Alonso-Matilla}},\ and\ \bibinfo {author} {\bibfnamefont {D.}~\bibnamefont {Saintillan}},\ }\bibfield  {title} {\bibinfo {title} {On the distribution and swim pressure of run-and-tumble particles in confinement},\ }\href {https://doi.org/10.1017/jfm.2015.520} {\bibfield  {journal} {\bibinfo  {journal} {Journal of Fluid Mechanics}\ }\textbf {\bibinfo {volume} {781}},\ \bibinfo {pages} {R4} (\bibinfo {year} {2015})}\BibitemShut {NoStop}%
\bibitem [{\citenamefont {Yan}\ and\ \citenamefont {Brady}(2015{\natexlab{a}})}]{Yan2015}%
  \BibitemOpen
  \bibfield  {author} {\bibinfo {author} {\bibfnamefont {W.}~\bibnamefont {Yan}}\ and\ \bibinfo {author} {\bibfnamefont {J.~F.}\ \bibnamefont {Brady}},\ }\bibfield  {title} {\bibinfo {title} {The force on a boundary in active matter},\ }\href {https://doi.org/10.1017/jfm.2015.621} {\bibfield  {journal} {\bibinfo  {journal} {Journal of Fluid Mechanics}\ }\textbf {\bibinfo {volume} {785}},\ \bibinfo {pages} {R1} (\bibinfo {year} {2015}{\natexlab{a}})}\BibitemShut {NoStop}%
\bibitem [{\citenamefont {Yan}\ and\ \citenamefont {Brady}(2018)}]{Yan2018}%
  \BibitemOpen
  \bibfield  {author} {\bibinfo {author} {\bibfnamefont {W.}~\bibnamefont {Yan}}\ and\ \bibinfo {author} {\bibfnamefont {J.~F.}\ \bibnamefont {Brady}},\ }\bibfield  {title} {\bibinfo {title} {The curved kinetic boundary layer of active matter},\ }\href {https://doi.org/10.1039/C7SM01643C} {\bibfield  {journal} {\bibinfo  {journal} {Soft Matter}\ }\textbf {\bibinfo {volume} {14}},\ \bibinfo {pages} {279} (\bibinfo {year} {2018})}\BibitemShut {NoStop}%
\bibitem [{\citenamefont {Duzgun}\ and\ \citenamefont {Selinger}(2018)}]{Duzgun2018}%
  \BibitemOpen
  \bibfield  {author} {\bibinfo {author} {\bibfnamefont {A.}~\bibnamefont {Duzgun}}\ and\ \bibinfo {author} {\bibfnamefont {J.~V.}\ \bibnamefont {Selinger}},\ }\bibfield  {title} {\bibinfo {title} {Active brownian particles near straight or curved walls: Pressure and boundary layers},\ }\href {https://doi.org/10.1103/PhysRevE.97.032606} {\bibfield  {journal} {\bibinfo  {journal} {Physical Review E}\ }\textbf {\bibinfo {volume} {97}},\ \bibinfo {pages} {032606} (\bibinfo {year} {2018})}\BibitemShut {NoStop}%
\bibitem [{\citenamefont {Granek}\ \emph {et~al.}(2024)\citenamefont {Granek}, \citenamefont {Kafri}, \citenamefont {Kardar}, \citenamefont {Ro}, \citenamefont {Tailleur},\ and\ \citenamefont {Solon}}]{Granek2024}%
  \BibitemOpen
  \bibfield  {author} {\bibinfo {author} {\bibfnamefont {O.}~\bibnamefont {Granek}}, \bibinfo {author} {\bibfnamefont {Y.}~\bibnamefont {Kafri}}, \bibinfo {author} {\bibfnamefont {M.}~\bibnamefont {Kardar}}, \bibinfo {author} {\bibfnamefont {S.}~\bibnamefont {Ro}}, \bibinfo {author} {\bibfnamefont {J.}~\bibnamefont {Tailleur}},\ and\ \bibinfo {author} {\bibfnamefont {A.}~\bibnamefont {Solon}},\ }\bibfield  {title} {\bibinfo {title} {Colloquium: Inclusions, boundaries, and disorder in scalar active matter},\ }\href {https://doi.org/10.1103/RevModPhys.96.031003} {\bibfield  {journal} {\bibinfo  {journal} {Reviews of Modern Physics}\ }\textbf {\bibinfo {volume} {96}},\ \bibinfo {pages} {031003} (\bibinfo {year} {2024})}\BibitemShut {NoStop}%
\bibitem [{\citenamefont {Nikola}\ \emph {et~al.}(2016)\citenamefont {Nikola}, \citenamefont {Solon}, \citenamefont {Kafri}, \citenamefont {Kardar}, \citenamefont {Tailleur},\ and\ \citenamefont {Voituriez}}]{Nikola2016}%
  \BibitemOpen
  \bibfield  {author} {\bibinfo {author} {\bibfnamefont {N.}~\bibnamefont {Nikola}}, \bibinfo {author} {\bibfnamefont {A.~P.}\ \bibnamefont {Solon}}, \bibinfo {author} {\bibfnamefont {Y.}~\bibnamefont {Kafri}}, \bibinfo {author} {\bibfnamefont {M.}~\bibnamefont {Kardar}}, \bibinfo {author} {\bibfnamefont {J.}~\bibnamefont {Tailleur}},\ and\ \bibinfo {author} {\bibfnamefont {R.}~\bibnamefont {Voituriez}},\ }\bibfield  {title} {\bibinfo {title} {Active particles with soft and curved walls: Equation of state, ratchets, and instabilities},\ }\href {https://doi.org/10.1103/PhysRevLett.117.098001} {\bibfield  {journal} {\bibinfo  {journal} {Physical Review Letters}\ }\textbf {\bibinfo {volume} {117}},\ \bibinfo {pages} {098001} (\bibinfo {year} {2016})}\BibitemShut {NoStop}%
\bibitem [{\citenamefont {Omar}\ \emph {et~al.}(2023)\citenamefont {Omar}, \citenamefont {Row}, \citenamefont {Mallory},\ and\ \citenamefont {Brady}}]{Omar2023}%
  \BibitemOpen
  \bibfield  {author} {\bibinfo {author} {\bibfnamefont {A.~K.}\ \bibnamefont {Omar}}, \bibinfo {author} {\bibfnamefont {H.}~\bibnamefont {Row}}, \bibinfo {author} {\bibfnamefont {S.~A.}\ \bibnamefont {Mallory}},\ and\ \bibinfo {author} {\bibfnamefont {J.~F.}\ \bibnamefont {Brady}},\ }\bibfield  {title} {\bibinfo {title} {Mechanical theory of nonequilibrium coexistence and motility-induced phase separation},\ }\href {https://pnas.org/doi/10.1073/pnas.2219900120} {\bibfield  {journal} {\bibinfo  {journal} {Proceedings of the National Academy of Sciences}\ }\textbf {\bibinfo {volume} {120}} (\bibinfo {year} {2023})}\BibitemShut {NoStop}%
\bibitem [{\citenamefont {Yan}\ and\ \citenamefont {Brady}(2015{\natexlab{b}})}]{Yan2015Swim}%
  \BibitemOpen
  \bibfield  {author} {\bibinfo {author} {\bibfnamefont {W.}~\bibnamefont {Yan}}\ and\ \bibinfo {author} {\bibfnamefont {J.~F.}\ \bibnamefont {Brady}},\ }\bibfield  {title} {\bibinfo {title} {The swim force as a body force},\ }\href {https://doi.org/10.1039/C5SM01318F} {\bibfield  {journal} {\bibinfo  {journal} {Soft Matter}\ }\textbf {\bibinfo {volume} {11}},\ \bibinfo {pages} {6235} (\bibinfo {year} {2015}{\natexlab{b}})}\BibitemShut {NoStop}%
\bibitem [{\citenamefont {Epstein}\ \emph {et~al.}(2019)\citenamefont {Epstein}, \citenamefont {Klymko},\ and\ \citenamefont {Mandadapu}}]{Epstein2019}%
  \BibitemOpen
  \bibfield  {author} {\bibinfo {author} {\bibfnamefont {J.~M.}\ \bibnamefont {Epstein}}, \bibinfo {author} {\bibfnamefont {K.}~\bibnamefont {Klymko}},\ and\ \bibinfo {author} {\bibfnamefont {K.~K.}\ \bibnamefont {Mandadapu}},\ }\bibfield  {title} {\bibinfo {title} {Statistical mechanics of transport processes in active fluids. ii. equations of hydrodynamics for active brownian particles},\ }\href {https://pubs.aip.org/jcp/article/150/16/164111/198165/Statistical-mechanics-of-transport-processes-in} {\bibfield  {journal} {\bibinfo  {journal} {The Journal of Chemical Physics}\ }\textbf {\bibinfo {volume} {150}} (\bibinfo {year} {2019})}\BibitemShut {NoStop}%
\bibitem [{\citenamefont {Omar}\ \emph {et~al.}(2020)\citenamefont {Omar}, \citenamefont {Wang},\ and\ \citenamefont {Brady}}]{Omar2020}%
  \BibitemOpen
  \bibfield  {author} {\bibinfo {author} {\bibfnamefont {A.~K.}\ \bibnamefont {Omar}}, \bibinfo {author} {\bibfnamefont {Z.-G.}\ \bibnamefont {Wang}},\ and\ \bibinfo {author} {\bibfnamefont {J.~F.}\ \bibnamefont {Brady}},\ }\bibfield  {title} {\bibinfo {title} {Microscopic origins of the swim pressure and the anomalous surface tension of active matter},\ }\href {https://doi.org/10.1103/PhysRevE.101.012604} {\bibfield  {journal} {\bibinfo  {journal} {Physical Review E}\ }\textbf {\bibinfo {volume} {101}},\ \bibinfo {pages} {012604} (\bibinfo {year} {2020})}\BibitemShut {NoStop}%
\bibitem [{Note3()}]{Note3}%
  \BibitemOpen
  \bibinfo {note} {We choose this name since $P_{\protect \rm c}$ originates from purely conservative forces, however its value is still dependent on activity~\cite {Solon2015, Omar2023, Speck2021, SeeDetails}}\BibitemShut {NoStop}%
\bibitem [{\citenamefont {Mallory}\ \emph {et~al.}(2021)\citenamefont {Mallory}, \citenamefont {Omar},\ and\ \citenamefont {Brady}}]{Mallory2021}%
  \BibitemOpen
  \bibfield  {author} {\bibinfo {author} {\bibfnamefont {S.~A.}\ \bibnamefont {Mallory}}, \bibinfo {author} {\bibfnamefont {A.~K.}\ \bibnamefont {Omar}},\ and\ \bibinfo {author} {\bibfnamefont {J.~F.}\ \bibnamefont {Brady}},\ }\bibfield  {title} {\bibinfo {title} {Dynamic overlap concentration scale of active colloids},\ }\href {https://doi.org/10.1103/PhysRevE.104.044612} {\bibfield  {journal} {\bibinfo  {journal} {Physical Review E}\ }\textbf {\bibinfo {volume} {104}},\ \bibinfo {pages} {044612} (\bibinfo {year} {2021})}\BibitemShut {NoStop}%
\bibitem [{\citenamefont {Takatori}\ \emph {et~al.}(2014)\citenamefont {Takatori}, \citenamefont {Yan},\ and\ \citenamefont {Brady}}]{Takatori2014}%
  \BibitemOpen
  \bibfield  {author} {\bibinfo {author} {\bibfnamefont {S.~C.}\ \bibnamefont {Takatori}}, \bibinfo {author} {\bibfnamefont {W.}~\bibnamefont {Yan}},\ and\ \bibinfo {author} {\bibfnamefont {J.~F.}\ \bibnamefont {Brady}},\ }\bibfield  {title} {\bibinfo {title} {Swim pressure: Stress generation in active matter},\ }\href {https://doi.org/10.1103/PhysRevLett.113.028103} {\bibfield  {journal} {\bibinfo  {journal} {Physical Review Letters}\ }\textbf {\bibinfo {volume} {113}},\ \bibinfo {pages} {028103} (\bibinfo {year} {2014})}\BibitemShut {NoStop}%
\bibitem [{\citenamefont {Cates}\ and\ \citenamefont {Tailleur}(2015)}]{Cates2015}%
  \BibitemOpen
  \bibfield  {author} {\bibinfo {author} {\bibfnamefont {M.~E.}\ \bibnamefont {Cates}}\ and\ \bibinfo {author} {\bibfnamefont {J.}~\bibnamefont {Tailleur}},\ }\bibfield  {title} {\bibinfo {title} {Motility-induced phase separation},\ }\href {https://doi.org/10.1146/annurev-conmatphys-031214-014710} {\bibfield  {journal} {\bibinfo  {journal} {Annual Review of Condensed Matter Physics}\ }\textbf {\bibinfo {volume} {6}},\ \bibinfo {pages} {219} (\bibinfo {year} {2015})}\BibitemShut {NoStop}%
\bibitem [{Note4()}]{Note4}%
  \BibitemOpen
  \bibinfo {note} {Similarly, the passive suspension ($\alpha =0$) must be stable, i.e. $\protect \dot {P}_{\protect \rm c}(\rho ) \geq 0$.}\BibitemShut {Stop}%
\bibitem [{Note5()}]{Note5}%
  \BibitemOpen
  \bibinfo {note} {Despite appearances, these are essential: derivatives holding $z$ and a component of $\protect \mathbf {x}$ constant are different from those holding $\xi $ and the corresponding component of $\protect \bm {\mu }$ fixed.}\BibitemShut {Stop}%
\bibitem [{\citenamefont {Bender}\ and\ \citenamefont {Orszag}(1978)}]{Bender1978}%
  \BibitemOpen
  \bibfield  {author} {\bibinfo {author} {\bibfnamefont {C.~M.}\ \bibnamefont {Bender}}\ and\ \bibinfo {author} {\bibfnamefont {S.~A.}\ \bibnamefont {Orszag}},\ }\href@noop {} {\emph {\bibinfo {title} {Advanced Mathematical Methods for Scientists and Engineers}}}\ (\bibinfo  {publisher} {McGraw-Hill Book Company},\ \bibinfo {year} {1978})\BibitemShut {NoStop}%
\bibitem [{\citenamefont {Dingle}(1973)}]{Dingle1973}%
  \BibitemOpen
  \bibfield  {author} {\bibinfo {author} {\bibfnamefont {R.~B.}\ \bibnamefont {Dingle}},\ }\href@noop {} {\emph {\bibinfo {title} {Asymptotic Expansions: Their Derivation and Interpretation}}}\ (\bibinfo  {publisher} {Academic Press},\ \bibinfo {year} {1973})\BibitemShut {NoStop}%
\bibitem [{See()}]{SeeDetails}%
  \BibitemOpen
  \href@noop {} {\bibinfo {title} {See supplemental material for details}}\BibitemShut {NoStop}%
\bibitem [{Note6()}]{Note6}%
  \BibitemOpen
  \bibinfo {note} {Note that the $q^0$ term is always absent: $c_1(q=0)$ exactly cancels the contribution from the second term in Eq.~\protect \eqref {eq:psiexp}.}\BibitemShut {Stop}%
\bibitem [{Note7()}]{Note7}%
  \BibitemOpen
  \bibinfo {note} {This classification is distinct from that of Cross-Hohenberg~\cite {Cross1993}, and essentially divides their ``stationary'' instabilities based on the fastest growing wavelength.}\BibitemShut {Stop}%
\bibitem [{\citenamefont {Maini}\ and\ \citenamefont {Woolley}(2019)}]{Maini2019}%
  \BibitemOpen
  \bibfield  {author} {\bibinfo {author} {\bibfnamefont {P.~K.}\ \bibnamefont {Maini}}\ and\ \bibinfo {author} {\bibfnamefont {T.~E.}\ \bibnamefont {Woolley}},\ }\bibinfo {title} {The turing model for biological pattern formation},\ in\ \href {https://doi.org/10.1007/978-3-030-22583-4_7} {\emph {\bibinfo {booktitle} {The Dynamics of Biological Systems.}}},\ \bibinfo {editor} {edited by\ \bibinfo {editor} {\bibfnamefont {A.}~\bibnamefont {Bianchi}}, \bibinfo {editor} {\bibfnamefont {T.}~\bibnamefont {Hillen}}, \bibinfo {editor} {\bibfnamefont {M.}~\bibnamefont {Lewis}},\ and\ \bibinfo {editor} {\bibfnamefont {Y.}~\bibnamefont {Yi}}}\ (\bibinfo  {publisher} {Springer, Cham},\ \bibinfo {year} {2019})\ pp.\ \bibinfo {pages} {189--204}\BibitemShut {NoStop}%
\bibitem [{\citenamefont {Yu}\ and\ \citenamefont {Košmrlj}(2025)}]{Yu2025}%
  \BibitemOpen
  \bibfield  {author} {\bibinfo {author} {\bibfnamefont {Q.}~\bibnamefont {Yu}}\ and\ \bibinfo {author} {\bibfnamefont {A.}~\bibnamefont {Košmrlj}},\ }\bibfield  {title} {\bibinfo {title} {Pattern formation of lipid domains in bilayer membranes},\ }\href {https://doi.org/10.1039/D5SM00276A} {\bibfield  {journal} {\bibinfo  {journal} {Soft Matter}\ }\textbf {\bibinfo {volume} {21}},\ \bibinfo {pages} {4288} (\bibinfo {year} {2025})}\BibitemShut {NoStop}%
\bibitem [{\citenamefont {Omar}\ \emph {et~al.}(2021)\citenamefont {Omar}, \citenamefont {Klymko}, \citenamefont {GrandPre},\ and\ \citenamefont {Geissler}}]{Omar2021}%
  \BibitemOpen
  \bibfield  {author} {\bibinfo {author} {\bibfnamefont {A.~K.}\ \bibnamefont {Omar}}, \bibinfo {author} {\bibfnamefont {K.}~\bibnamefont {Klymko}}, \bibinfo {author} {\bibfnamefont {T.}~\bibnamefont {GrandPre}},\ and\ \bibinfo {author} {\bibfnamefont {P.~L.}\ \bibnamefont {Geissler}},\ }\bibfield  {title} {\bibinfo {title} {Phase diagram of active brownian spheres: Crystallization and the metastability of motility-induced phase separation},\ }\href {https://doi.org/10.1103/PhysRevLett.126.188002} {\bibfield  {journal} {\bibinfo  {journal} {Physical Review Letters}\ }\textbf {\bibinfo {volume} {126}},\ \bibinfo {pages} {188002} (\bibinfo {year} {2021})}\BibitemShut {NoStop}%
\bibitem [{\citenamefont {Bergert}\ \emph {et~al.}(2021)\citenamefont {Bergert}, \citenamefont {Lembo}, \citenamefont {Sharma}, \citenamefont {Russo}, \citenamefont {Milovanović}, \citenamefont {Gretarsson}, \citenamefont {Börmel}, \citenamefont {Neveu}, \citenamefont {Hackett}, \citenamefont {Petsalaki},\ and\ \citenamefont {Diz-Muñoz}}]{Bergert2021}%
  \BibitemOpen
  \bibfield  {author} {\bibinfo {author} {\bibfnamefont {M.}~\bibnamefont {Bergert}}, \bibinfo {author} {\bibfnamefont {S.}~\bibnamefont {Lembo}}, \bibinfo {author} {\bibfnamefont {S.}~\bibnamefont {Sharma}}, \bibinfo {author} {\bibfnamefont {L.}~\bibnamefont {Russo}}, \bibinfo {author} {\bibfnamefont {D.}~\bibnamefont {Milovanović}}, \bibinfo {author} {\bibfnamefont {K.~H.}\ \bibnamefont {Gretarsson}}, \bibinfo {author} {\bibfnamefont {M.}~\bibnamefont {Börmel}}, \bibinfo {author} {\bibfnamefont {P.~A.}\ \bibnamefont {Neveu}}, \bibinfo {author} {\bibfnamefont {J.~A.}\ \bibnamefont {Hackett}}, \bibinfo {author} {\bibfnamefont {E.}~\bibnamefont {Petsalaki}},\ and\ \bibinfo {author} {\bibfnamefont {A.}~\bibnamefont {Diz-Muñoz}},\ }\bibfield  {title} {\bibinfo {title} {Cell surface mechanics gate embryonic stem cell differentiation},\ }\href {https://doi.org/10.1016/j.stem.2020.10.017} {\bibfield  {journal} {\bibinfo  {journal} {Cell Stem Cell}\ }\textbf {\bibinfo {volume} {28}},\ \bibinfo {pages} {209}
  (\bibinfo {year} {2021})}\BibitemShut {NoStop}%
\bibitem [{\citenamefont {Belly}\ \emph {et~al.}(2021)\citenamefont {Belly}, \citenamefont {Stubb}, \citenamefont {Yanagida}, \citenamefont {Labouesse}, \citenamefont {Jones}, \citenamefont {Paluch},\ and\ \citenamefont {Chalut}}]{DeBelly2021}%
  \BibitemOpen
  \bibfield  {author} {\bibinfo {author} {\bibfnamefont {H.~D.}\ \bibnamefont {Belly}}, \bibinfo {author} {\bibfnamefont {A.}~\bibnamefont {Stubb}}, \bibinfo {author} {\bibfnamefont {A.}~\bibnamefont {Yanagida}}, \bibinfo {author} {\bibfnamefont {C.}~\bibnamefont {Labouesse}}, \bibinfo {author} {\bibfnamefont {P.~H.}\ \bibnamefont {Jones}}, \bibinfo {author} {\bibfnamefont {E.~K.}\ \bibnamefont {Paluch}},\ and\ \bibinfo {author} {\bibfnamefont {K.~J.}\ \bibnamefont {Chalut}},\ }\bibfield  {title} {\bibinfo {title} {Membrane tension gates erk-mediated regulation of pluripotent cell fate},\ }\href {https://doi.org/10.1016/j.stem.2020.10.018} {\bibfield  {journal} {\bibinfo  {journal} {Cell Stem Cell}\ }\textbf {\bibinfo {volume} {28}},\ \bibinfo {pages} {273} (\bibinfo {year} {2021})}\BibitemShut {NoStop}%
\bibitem [{\citenamefont {Zhou}\ and\ \citenamefont {Brady}(2026)}]{Zhou2026}%
  \BibitemOpen
  \bibfield  {author} {\bibinfo {author} {\bibfnamefont {T.}~\bibnamefont {Zhou}}\ and\ \bibinfo {author} {\bibfnamefont {J.~F.}\ \bibnamefont {Brady}},\ }\bibfield  {title} {\bibinfo {title} {Hydrodynamic interactions destroy motility-induced phase separation in active suspensions},\ }\href {https://doi.org/10.1103/jhhk-j5g9} {\bibfield  {journal} {\bibinfo  {journal} {Physical Review Letters}\ }\textbf {\bibinfo {volume} {136}},\ \bibinfo {pages} {088301} (\bibinfo {year} {2026})}\BibitemShut {NoStop}%
\bibitem [{\citenamefont {Edwards}\ and\ \citenamefont {Freed}(1974)}]{Edwards1974}%
  \BibitemOpen
  \bibfield  {author} {\bibinfo {author} {\bibfnamefont {S.~F.}\ \bibnamefont {Edwards}}\ and\ \bibinfo {author} {\bibfnamefont {K.~F.}\ \bibnamefont {Freed}},\ }\bibfield  {title} {\bibinfo {title} {Theory of the dynamical viscosity of polymer solutions},\ }\href {https://doi.org/10.1063/1.1681993} {\bibfield  {journal} {\bibinfo  {journal} {The Journal of Chemical Physics}\ }\textbf {\bibinfo {volume} {61}},\ \bibinfo {pages} {1189} (\bibinfo {year} {1974})}\BibitemShut {NoStop}%
\bibitem [{\citenamefont {Freed}\ and\ \citenamefont {Edwards}(1974)}]{Freed1974}%
  \BibitemOpen
  \bibfield  {author} {\bibinfo {author} {\bibfnamefont {K.~F.}\ \bibnamefont {Freed}}\ and\ \bibinfo {author} {\bibfnamefont {S.~F.}\ \bibnamefont {Edwards}},\ }\bibfield  {title} {\bibinfo {title} {Polymer viscosity in concentrated solutions},\ }\href {https://doi.org/10.1063/1.1682545} {\bibfield  {journal} {\bibinfo  {journal} {The Journal of Chemical Physics}\ }\textbf {\bibinfo {volume} {61}},\ \bibinfo {pages} {3626} (\bibinfo {year} {1974})}\BibitemShut {NoStop}%
\bibitem [{\citenamefont {Shaqfeh}\ and\ \citenamefont {Fredrickson}(1990)}]{Shaqfeh1990}%
  \BibitemOpen
  \bibfield  {author} {\bibinfo {author} {\bibfnamefont {E.~S.~G.}\ \bibnamefont {Shaqfeh}}\ and\ \bibinfo {author} {\bibfnamefont {G.~H.}\ \bibnamefont {Fredrickson}},\ }\bibfield  {title} {\bibinfo {title} {The hydrodynamic stress in a suspension of rods},\ }\href {https://doi.org/10.1063/1.857683} {\bibfield  {journal} {\bibinfo  {journal} {Physics of Fluids A: Fluid Dynamics}\ }\textbf {\bibinfo {volume} {2}},\ \bibinfo {pages} {7} (\bibinfo {year} {1990})}\BibitemShut {NoStop}%
\bibitem [{\citenamefont {Henderson}(1979)}]{Henderson1979}%
  \BibitemOpen
  \bibfield  {author} {\bibinfo {author} {\bibfnamefont {D.}~\bibnamefont {Henderson}},\ }\bibfield  {title} {\bibinfo {title} {An exact formula for the contact value of the density profile of a system of charged hard spheres near a charged wall},\ }\href {https://doi.org/10.1016/0368-1874(79)87191-9} {\bibfield  {journal} {\bibinfo  {journal} {Journal of Electroanalytical Chemistry}\ }\textbf {\bibinfo {volume} {102}},\ \bibinfo {pages} {315} (\bibinfo {year} {1979})}\BibitemShut {NoStop}%
\bibitem [{\citenamefont {Solon}\ \emph {et~al.}(2015)\citenamefont {Solon}, \citenamefont {Stenhammar}, \citenamefont {Wittkowski}, \citenamefont {Kardar}, \citenamefont {Kafri}, \citenamefont {Cates},\ and\ \citenamefont {Tailleur}}]{Solon2015}%
  \BibitemOpen
  \bibfield  {author} {\bibinfo {author} {\bibfnamefont {A.~P.}\ \bibnamefont {Solon}}, \bibinfo {author} {\bibfnamefont {J.}~\bibnamefont {Stenhammar}}, \bibinfo {author} {\bibfnamefont {R.}~\bibnamefont {Wittkowski}}, \bibinfo {author} {\bibfnamefont {M.}~\bibnamefont {Kardar}}, \bibinfo {author} {\bibfnamefont {Y.}~\bibnamefont {Kafri}}, \bibinfo {author} {\bibfnamefont {M.~E.}\ \bibnamefont {Cates}},\ and\ \bibinfo {author} {\bibfnamefont {J.}~\bibnamefont {Tailleur}},\ }\bibfield  {title} {\bibinfo {title} {Pressure and phase equilibria in interacting active brownian spheres},\ }\href {https://doi.org/10.1103/PhysRevLett.114.198301} {\bibfield  {journal} {\bibinfo  {journal} {Physical Review Letters}\ }\textbf {\bibinfo {volume} {114}},\ \bibinfo {pages} {198301} (\bibinfo {year} {2015})}\BibitemShut {NoStop}%
\bibitem [{\citenamefont {Speck}(2021)}]{Speck2021}%
  \BibitemOpen
  \bibfield  {author} {\bibinfo {author} {\bibfnamefont {T.}~\bibnamefont {Speck}},\ }\bibfield  {title} {\bibinfo {title} {Coexistence of active brownian disks: van der waals theory and analytical results},\ }\href {https://doi.org/10.1103/PhysRevE.103.012607} {\bibfield  {journal} {\bibinfo  {journal} {Physical Review E}\ }\textbf {\bibinfo {volume} {103}},\ \bibinfo {pages} {012607} (\bibinfo {year} {2021})}\BibitemShut {NoStop}%
\bibitem [{\citenamefont {Cross}\ and\ \citenamefont {Hohenberg}(1993)}]{Cross1993}%
  \BibitemOpen
  \bibfield  {author} {\bibinfo {author} {\bibfnamefont {M.~C.}\ \bibnamefont {Cross}}\ and\ \bibinfo {author} {\bibfnamefont {P.~C.}\ \bibnamefont {Hohenberg}},\ }\bibfield  {title} {\bibinfo {title} {Pattern formation outside of equilibrium},\ }\href {https://doi.org/10.1103/RevModPhys.65.851} {\bibfield  {journal} {\bibinfo  {journal} {Reviews of Modern Physics}\ }\textbf {\bibinfo {volume} {65}},\ \bibinfo {pages} {851} (\bibinfo {year} {1993})}\BibitemShut {NoStop}%
\bibitem [{\citenamefont {Aifantis}\ and\ \citenamefont {Serrin}(1983{\natexlab{a}})}]{Aifantis1983}%
  \BibitemOpen
  \bibfield  {author} {\bibinfo {author} {\bibfnamefont {E.~C.}\ \bibnamefont {Aifantis}}\ and\ \bibinfo {author} {\bibfnamefont {J.~B.}\ \bibnamefont {Serrin}},\ }\bibfield  {title} {\bibinfo {title} {Equilibrium solutions in the mechanical theory of fluid microstructures},\ }\href {https://doi.org/10.1016/0021-9797(83)90054-1} {\bibfield  {journal} {\bibinfo  {journal} {Journal of Colloid and Interface Science}\ }\textbf {\bibinfo {volume} {96}},\ \bibinfo {pages} {530} (\bibinfo {year} {1983}{\natexlab{a}})}\BibitemShut {NoStop}%
\bibitem [{\citenamefont {Aifantis}\ and\ \citenamefont {Serrin}(1983{\natexlab{b}})}]{Aifantis1983Maxwell}%
  \BibitemOpen
  \bibfield  {author} {\bibinfo {author} {\bibfnamefont {E.~C.}\ \bibnamefont {Aifantis}}\ and\ \bibinfo {author} {\bibfnamefont {J.~B.}\ \bibnamefont {Serrin}},\ }\bibfield  {title} {\bibinfo {title} {The mechanical theory of fluid interfaces and maxwell's rule},\ }\href {https://doi.org/10.1016/0021-9797(83)90053-X} {\bibfield  {journal} {\bibinfo  {journal} {Journal of Colloid and Interface Science}\ }\textbf {\bibinfo {volume} {96}},\ \bibinfo {pages} {517} (\bibinfo {year} {1983}{\natexlab{b}})}\BibitemShut {NoStop}%
\end{thebibliography}
%

	\appendix
    \section{Separation of Timescales}
    \label{app:Times}
    Consider a perturbation to a uniform density and polarization state \textit{along the membrane}, i.e. $\rho = \rho_0 +\delta\rho(\mathbf{x},t)$ and $\mathbf{F} = \mathbf{F}_0 + \bm{\delta F}(\mathbf{x},t)$. 
    Linearizing Eqns.~\eqref{eq:ctyrho} in the perturbations and assuming they relax $\sim e^{-t/\tau_{\rho}}$, we find (in FT):
    \begin{equation}
      \frac{T_d}{\tau_\rho}\begin{pmatrix}
    \widetilde{\delta\rho} \\
    \widetilde{\bm{\delta F}}_{\parallel}
\end{pmatrix}
 =  \begin{pmatrix}
K_{\rm c}\Lambda_d^2 q^2/k_B T &  i \mathbf{q} \Lambda_d^2/k_B T\\
 i \mathbf{q} K_{\rm act} & (\Lambda_d^2 q^2 +1)
\end{pmatrix} 
\begin{pmatrix}
    \widetilde{\delta\rho} \\
    \widetilde{\bm{\delta F}}_{\parallel}
\end{pmatrix}
    \end{equation}
    where $T_d = \tau_r/(d-1)$, $K_{\rm c/act} = \dot{P}_{\rm c/act}(\rho_0)$ and $\bm{\delta F}_{\parallel}$ is the component of the polarization perturbation parallel to the membrane (it is easy to show that the perpendicular component relaxes quickly, at a rate $= (\Lambda_d^2 q^2 +1) T_d^{-1}$). 
    The density relaxation rate is therefore controlled by the \textit{smallest} eigenvalue of the matrix on the right hand side. 
    In the limit $\Lambda_d q \ll 1$, this is
    \begin{equation}
        T_d \tau_{\rho}^{-1} \approx (K_{\rm c}+ K_{\rm act}) (\Lambda_d q)^2/k_BT.  
    \end{equation}
    Throughout this work, we consider the case that MIPS does not occur so that the first bracket on the right hand side is always positive. 
    MIPS occurs at high densities and activities, thus we should restrict our attention to relatively low densities where this bracket is $\approx (1+\alpha^2)$.
    
    The relaxation rate, $\tau_h^{-1}$, of a membrane perturbation with wavenumber $q$ is found from Eq.~\eqref{eq:EqnMot} in $d=3$ to be: 
    \begin{equation}
        T_d \tau_h^{-1} = 3 g (\Lambda_d q) + 3 k (\Lambda_d q)^3/4.
    \end{equation}
    For the adiabatic approximation to hold, the ABP density must relax faster than $h$, i.e. $\tau_{\rho}^{-1} > \tau_{h}^{-1}$.
    Thus, we conclude that the adiabatic approximation can only be accurate if $\alpha^2 \gtrsim 3 \sqrt{g k}  - 1$ and it describes wavenumbers $g/(1+\alpha^2)\lesssim\Lambda_d q \lesssim (1+\alpha^2)/k$.
    The first condition means $\alpha \gg 1$, for typical membrane properties, $10^0\lesssim g \lesssim 10^1 $ and $10^1\lesssim k \lesssim 10^2$. 
    This gives the rough approximations quoted in the main text. 
    
    \section{Solution Details}
    \label{app:solcon}
    \textit{Coordinate Transformation:} In the $(\bm{\mu},\xi)$ coordinates introduced in the main text, after applying the multivariable chain rule, the gradient of a scalar function is
    \begin{equation}
    \label{eq:appgeo}
        \bm{\nabla} f = \partial_{\xi}f \ \hat{\mathbf{z}} + \bm{\nabla}_{\mu} f - \bm{\nabla}_{\mu} h \ \partial_{\xi} f,
    \end{equation}
    where $\bm{\nabla}_{\mu} f = (\partial_{\mathbf{x}}f)_{\xi}$, is the gradient along the membrane while keeping $\xi$ constant, \textit{rather than} $z$. 
    We use this to write Eqs.~\eqref{eq:sseqns} to linear order in $\bm{\nabla} h$ before taking the expansions in Eqs.~\eqref{eq:appexp}.

\textit{Linking conservative and active pressure corrections:} The first order corrections to the conservative and active pressures are linked by their shared dependence on the density profile, $\rho(\bm{\mu},\xi)$. 
Let $\rho_{0}(\xi)$ be the flat wall density profile and $\rho_1(\bm{\mu},\xi)$ its first order correction. 
The conservative (or active) pressure outside the deformed membrane must be: $P_{\rm c}(\rho_0+\rho_{1}) = P_c(\rho_0) + \dot{P_c}(\rho_0) \rho_1$, where a dot is a $\rho$ derivative. 
The first term here is, by definition, $P_{\rm c}^{(0)}$ and the second includes $p_c$ (along with the second term on the left hand side of Eq.~\eqref{eq:Piexp}). 
Expanding $P_{\rm act}$ as above, gives $\rho_1$ in terms of \textit{either} $p_{\rm c}$ or $p_{\rm act}$.
Both must be equal, and we obtain Eq.~\eqref{eq:chi}.

\textit{Flat wall solutions:} Let us provide some useful general results for the flat wall solutions. 
    It is straightforward to show that, $F^{(0)}(\xi) = \partial_{\xi}P_{\rm c}^{(0)}$ and hence
    \begin{equation}
        \Lambda_d^2 \partial_{\xi}^{2} P_c^{(0)} - \left(P_{\rm c}^{(0)} +P_{\rm act}^{(0)} - P_c^{(0)}(0)\right) = 0. 
    \end{equation}    
This equation, paired with the boundary conditions ${P_c(\xi \to \infty) = P_c(\rho_{\infty})}$ and $\lim_{\xi \to 0}(\Lambda_d^2\partial_{\xi}^{2}P_c^{(0)}-P_{\rm act}^{(0)})=0$ allows $P_c^{(0)}$ to be found as a series near $\xi = 0$ easily, under the assumptions of no MIPS and repulsive interactions (i.e. $\dot{P}_{\rm c}(\rho) > 0$). 
For our purposes the first two terms, $P_{\rm c}^{0} = P_{\rm c}^{(0)}(0) +\xi P_{\rm c}^{(1)} +\cdots$, are sufficient and the coefficients are given by:
\begin{subequations}
\label{eq:appflat}
    \begin{equation}
        P_{\rm c}^{(0)}(0) = P_{\rm c}(\rho_{\infty}) + P_{\rm act}(\rho_{\infty}) \equiv P_c(\rho_0(0)),
    \end{equation}
    \begin{equation}
        \frac{1}{2}\left(\Lambda_d P_c^{(1)}\right)^2 = \int_{P_{\rm c}(\rho_{\infty})}^{P_{\rm c}^{(0)}(0)} d P_c \left(P_{\rm c} + P_{\rm act}(P_{\rm c}) - P_{\rm c}^{(0)}(0)\right).
    \end{equation}
\end{subequations}
The first of these is well-known~\cite{Duzgun2018,Omar2020,Takatori2014} and establishes the density at a flat membrane $\rho_0(0)$ while the second may be found using a simple modification of the method of Aifantis and Serrin~\cite{Aifantis1983,Aifantis1983Maxwell,Omar2023}.

\textit{First order correction construction:} We find that $p_{\rm c}$ and $\bm{f}$ solve (in FT):
    \begin{subequations}
    \begin{equation}
        \partial_{\xi}^2 \tilde{p}_c - q^2 \tilde{p}_c  = \partial_{\xi} \tilde{f}_z + i \mathbf{q} \cdot \tilde{\bm{f}}_{\parallel},
    \end{equation}
    \begin{equation}
        \Lambda_d^2 \partial_{\xi}^2 \tilde{f}_z - (1+\Lambda_d^2 q^2) \tilde{f}_z  = \partial_{\xi} (\chi(\xi) \tilde{p_c}),
    \end{equation}
        \begin{equation}
        \Lambda_d^2 \partial_{\xi}^2 \tilde{\bm{f}}_{\parallel}- (1+\Lambda_d^2 q^2) \tilde{\bm{f}}_{\parallel}  = i \mathbf{q} \chi(\xi) \tilde{p_c},
    \end{equation}
    \end{subequations}
where $f_z$ and $\bm{f}_{\parallel}$ are the components of the polarization perpendicular and parallel to the flat membrane respectively. 
We construct the solutions for $\widetilde{p}_{c}$ and $\widetilde{\bm{f}}$ that vanish as $\xi \to \infty$ as follows:
	\begin{equation}
    \label{eq:pcapp}
		\widetilde{p}_{\rm c}(\mathbf{q},\xi) = c_1 \psi(\xi) + c_2 \varphi(\xi),
	\end{equation}
	where $\psi$ and $\varphi$ solve,
	\begin{subequations}
    \label{eq:psiphi}
		\begin{equation}
			\Lambda_d^2 \psi''(\xi) - (1 + \Lambda_d^2 q^2 +\chi(\xi)) \psi(\xi) = 0, 
		\end{equation} 
		subject to ${\psi(\xi=0)=1}$ and ${\psi(\xi \to \infty)=0}$ and, 
		\begin{equation}
        \label{eq:appGreen}
			\Lambda_d^2 \varphi''(\xi) - (1 + \Lambda_d^2 q^2 +\chi(\xi)) \varphi(\xi) = e^{-q \xi},
		\end{equation}
	\end{subequations}
	such that $\varphi(\xi=0)=0$ and ${\varphi(\xi \to \infty) \to 0}$. 
	\begin{equation}
		\begin{split}
			&\widetilde{\bm{f}}_{\parallel}(\mathbf{q},\xi) \equiv i \mathbf{q}\widetilde{f}_{\parallel}(\mathbf{q},\xi)\\
			& = i \mathbf{q}\Big[\widetilde{p}_{\rm c}(\mathbf{q},\xi) + c_2 e^{-q \xi}- c_3  q^{-2}\lambda(q) e^{- \lambda(q)\xi} \Big],
		\end{split}
	\end{equation}
	\begin{equation}
		\widetilde{f}_{z}(\mathbf{q},\xi) = \widetilde{f}_{\parallel}^{\prime}(\mathbf{q},\xi) - c_3 (\Lambda_d q)^{-2}e^{-\lambda(q)\xi},
	\end{equation}
    where $\lambda(q) = \sqrt{\Lambda_d^{-2}+ q^2}$.
	
	\textit{Boundary conditions:} The constants, $c_1$, $c_2$ and $c_3$, are found by matching the BCs in Eq.~\eqref{eq:BCs} at $\xi=0$. 
	\begin{subequations}
		\label{eq:CFBCSimp}
		\begin{equation}
			q c_2 - c_3 =0,
		\end{equation}
		\begin{equation}
			\lambda^2(q)(c_1 +c_2) - \lambda(q)c_3 = -  \Lambda_d^{-2}h(\mathbf{q},t) P_c^{(1)},
		\end{equation}
		\begin{equation}
			\mathcal{J}(q^2) c_1 + \left[\mathcal{K}(q)-q\right]c_2 + c_3 (\lambda(q)/q)^2 = -\Lambda_d^{-2}h(\mathbf{q},t) P_{\rm act}^{\rm 0}.
		\end{equation}
	\end{subequations}
	Here, we have defined ${P_{\rm act}^{\rm 0} \equiv P_{\rm act}(\rho_0(0))}$, as well as ${\mathcal{J}(q^2) = \psi'(\xi =0)}$ and ${\mathcal{K}(q) = \varphi'(\xi=0)}$.
    
    \textit{Non-interacting solution:} In the non-interacting limit, $\chi = \alpha^2$, and we find
    \begin{equation}
    \label{eq:appnonint}
        \psi(\xi) = e^{- \lambda_{\alpha}(q)\xi}, \ \ \ \text{and} \ \ \  \varphi(\xi) = \frac{e^{- \lambda_{\alpha}(q)\xi} - e^{- q \xi}}{1+\alpha^2},
    \end{equation}
    where we have defined $\lambda_{\alpha}(q) = \sqrt{\lambda^2(q)+ (\alpha/\Lambda_d)^2}$. 
    Using the well-known flat wall solutions~\cite{Yan2015} for non-interacting ABPs, we find the coefficients of Eq.~\eqref{eq:pext}
    \begin{subequations}
    \begin{equation}
     X_2 = - k_B T \Lambda_d\rho_{\infty} \alpha^2 \sqrt{1+\alpha^2},
    \end{equation}
    \begin{equation}
     X_3 = k_B T \Lambda_d^2 \rho_{\infty} \alpha^2 (1+2\alpha^2)/2, 
    \end{equation}
    \begin{equation}
     X_4 = - \frac{k_B T \Lambda_d^3\rho_{\infty} \alpha^2}{2(1+\alpha^2)} \left(d_1(\alpha) +d_2(\alpha)\sqrt{1+\alpha^2}\right), 
    \end{equation}
    \end{subequations}
    where ${d_1(\alpha)=2\alpha^4 + 3\alpha^2 + 1}$ and ${d_2(\alpha)=2\alpha^4 - \alpha^2 - 2}$.
    These expressions provide expressions for $\gamma_{\rm eff}$ and $\kappa_{\rm eff}$.

    \textit{Interacting Solution:} In the interacting case, we use WKB approximations for $\psi$ and $\varphi$~\cite{Bender1978,Dingle1973}:
    \begin{subequations}
        \begin{equation}
\label{eq:fWKB}
	\psi(\xi) \approx \left(\frac{Q(\mathbf{q},0)}{Q(\mathbf{q},\xi)}\right)^{1/4} \exp\left(- \frac{1}{\Lambda_d}\int_{0}^{\xi} d\xi' \sqrt{Q(\mathbf{q},\xi')}\right),
\end{equation}
\begin{equation}
    \varphi(\xi) = \int_{0}^{\infty} d \xi' G_Q(\xi,\xi') e^{-q\xi'},
\end{equation}
    \end{subequations}
    where $Q(\mathbf{q},\xi) = (1+\Lambda_d^2 q^2 + \chi(\xi))$ and $G_Q$ is the required Green function for Eq.~\eqref{eq:appGreen} (see Sec. 10.3 of~\cite{Bender1978}). 
    Since $\varphi$ is expressed as an integral, $\mathcal{K}(q)$ shall be also, which can be approximated as a series in powers of derivatives of $\chi$~\cite{SeeDetails}. 
    Concrete expressions for $X_{1\to4}$ are found in the Supplemental Material~\cite{SeeDetails}.
    
    \section{Changing membrane properties}
    \label{app:MemProp}
\begin{figure}[H]\includegraphics[width=8.8cm]{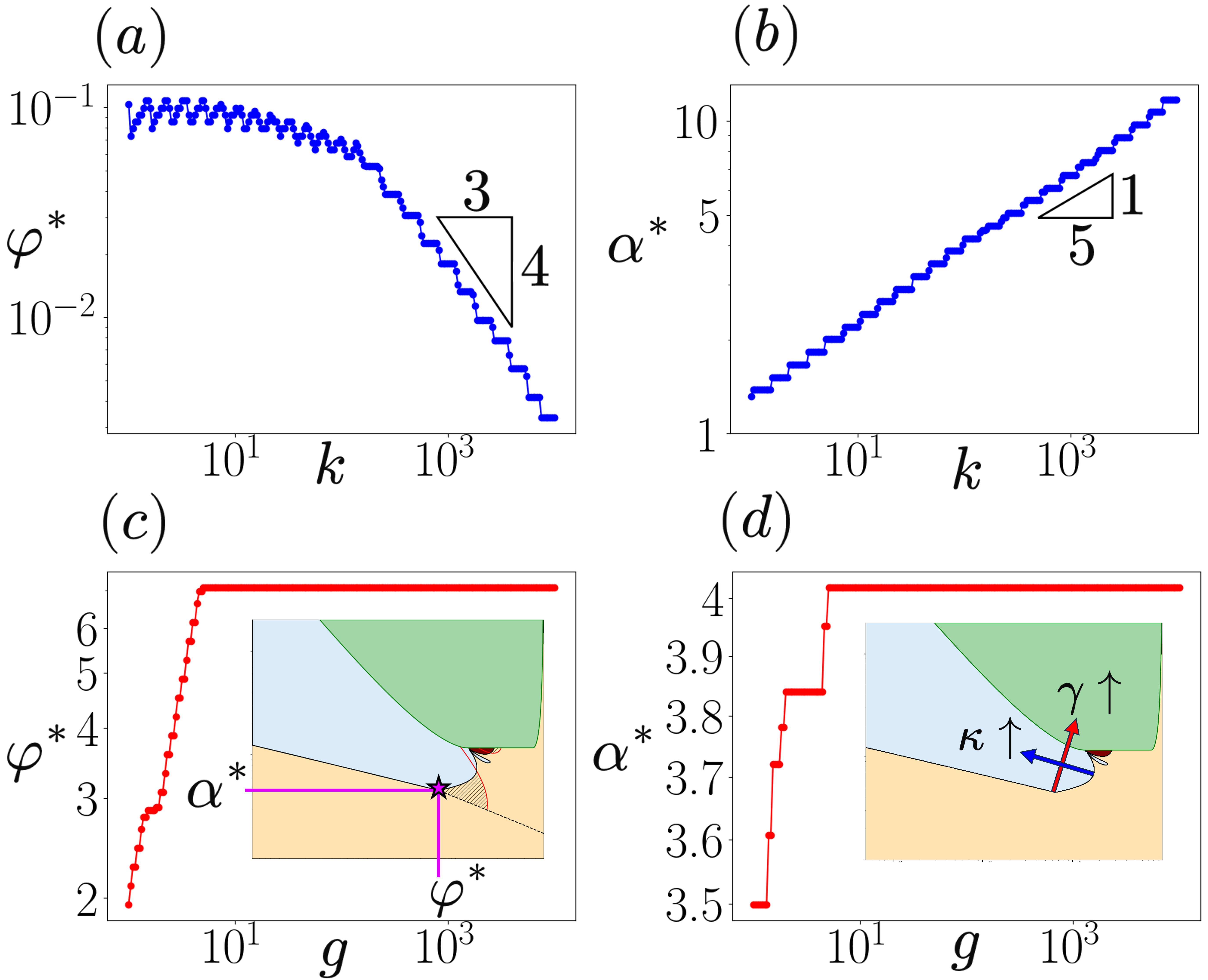}
	\caption{\label{fig:app} 
    Plots of the reduced volume fraction, $\varphi^{*}$ [(a) and (c)], and activity, $\alpha^{*}$ [(b) and (d)], where the interacting solution departs from the non-interacting solution [pink star, inset panel (c)] as functions of reduced bending modulus, $k$ panels (a) and (b), and surface tension, $g$ panels (c) and (d). Approximate scalings with $k$ are indicated. When held fixed, membrane parameters are $g=6$ and $k=154$, consistent with those in Fig.~\ref{fig:Fig}. These show how the blue region of short wavelength instability shrinks when $\gamma$ or $\kappa$ are increased, as shown by the red and blue arrows in the inset of panel (d). 
    }
\end{figure}

    Figure~\ref{fig:app} illustrates the dependencies of key features of the stability diagram on the membrane properties. \\

\end{document}


\title{Supplemental Material: Active Particles Destabilize Passive Membranes}

	\author{David A. King}
	\email{daking2@lbl.gov}
	\affiliation{Materials Science Division, Lawrence Berkeley National Laboratory, Berkeley CA 94720.}
	\affiliation{Department of Materials Science and Engineering, University of California, Berkeley, CA 94720.}
    
    \author{Thomas P. Russell}
    \affiliation{Polymer Science and Engineering Department, University of Massachusetts, Amherst MA 01003.}
    \affiliation{Materials Science Division, Lawrence Berkeley National Laboratory, Berkeley CA 94720.}
    
	\author{Ahmad K. Omar}
    \email{aomar@berkeley.edu}
	\affiliation{Materials Science Division, Lawrence Berkeley National Laboratory, Berkeley CA 94720.}
	\affiliation{Department of Materials Science and Engineering, University of California, Berkeley, CA 94720.}
    
\maketitle
\onecolumngrid
\tableofcontents

\section{\texorpdfstring{Expressions for $X_{1 \to 4}$ with interactions}{Expressions for X1-4 with interactions}}

In this section, we provide the expressions for the coefficients $X_{1 \to 4}$ appearing in Eq.~(10) of the main text. 
The non-interacting case was handled in the main appendix. 
Here we focus on the interacting case, and provide some additional details about their derivation. 
Our starting points are Eqs.~(B9) of the main text:
\begin{subequations}
	\label{eq:CFBCSimp}
		\begin{equation}
			q c_2(\mathbf{q}) - c_3(\mathbf{q}) =0,
		\end{equation}
		\begin{equation}
			\lambda^2(q)(c_1(\mathbf{q}) +c_2(\mathbf{q})) - \lambda(q)c_3(\mathbf{q}) = -  \Lambda_d^{-2}h(\mathbf{q},t) P_c^{(1)},
		\end{equation}
		\begin{equation}
			\mathcal{J}(q^2) c_1(\mathbf{q}) + \left[\mathcal{K}(q)-q\right]c_2(\mathbf{q}) + c_3(\mathbf{q}) (\lambda(q)/q)^2 = -\Lambda_d^{-2}h(\mathbf{q},t) P_{\rm act}^{\rm 0}.
		\end{equation}
\end{subequations}
Here, we have defined $\lambda(q) = \sqrt{\Lambda_d^{-2}+ q^2}$, ${P_{\rm act}^{\rm 0} \equiv P_{\rm act}(\rho_0(0))}$, as well as ${\mathcal{J}(q^2) = \psi'(\xi =0)}$ and ${\mathcal{K}(q) = \varphi'(\xi=0)}$.
Recall that primes denote differentiation with respect to $\xi$.
From these expressions we find $c_1(\mathbf{q})$ which is then expanded to fourth order in $q$ as: 
\begin{equation}
    c_1(\mathbf{q}) \approx X_1 q + X_2 q^2 + X_3 q^3 + X_4 q^4 + \cdots
\end{equation}
In order to find $X_{1 \to 4}$, therefore, we need expressions for $\mathcal{J}(q^2)$ and $\mathcal{K}(q)$. Let us handle each in turn. 

\subsection{\texorpdfstring{Approximating $\mathcal{J}(q^2)$}{Approximating J(q^2)}}
By definition $\mathcal{J}(q^2) = \psi'(\xi =0)$ and Eq.~(9) of the main text shows that $\psi$ solves
\begin{equation}
    \Lambda_d^2 \psi''(\xi) - \left(1 + \Lambda_d^2 q^2 + \chi(\xi)\right)\psi(\xi) = 0
\end{equation}
with boundary conditions $\psi(0)=1$ and $\psi(\infty) = 0$.
As discussed in the main text, at moderate activities and high densities, the solution can be approximated using WKB theory~\footnote{Although, this naming convention does not appropriately apportion the credit for its discovery. For a full discussion, see~\cite{Dingle1973}}.
\begin{equation}
\label{eq:fWKB}
	\psi(\xi) \approx \left(\frac{Q(\mathbf{q},0)}{Q(\mathbf{q},\xi)}\right)^{1/4} \exp\left(- \frac{1}{\Lambda_d}\int_{0}^{\xi} d\xi' \sqrt{Q(\mathbf{q},\xi')}\right),
\end{equation}
where $Q(\mathbf{q},\xi) = (1+\Lambda_d^2 q^2 + \chi(\xi))$.
Direct differentiation of this yields
\begin{equation}
\label{eq:Japprox}
    \mathcal{J}(q^2) \approx -\sqrt{Q(0)} \left(1 + \frac{Q'(0)}{4 Q^{3/2}(0)}\right) = -\sqrt{1+q^2 +\chi(0)} \left(1 + \frac{\chi'(0)}{4 (1+q^2 +\chi(0))^{3/2}}\right). 
\end{equation}
Notice that $\chi'(0)$ appears in this expression. 
It is not immediately obvious that this can be expressed in terms of quantities introduced in the main text and appendices, but a few applications of the chain rule clears this up. 
\begin{equation}
    \chi'(0) = \lim_{\xi \to 0} \frac{d}{d\xi} \frac{\dot{P}_{\rm act }(\rho_0(\xi))}{\dot{P}_{\rm c}(\rho_0(\xi))} = \frac{\dot{P}_{\rm act}(\rho_0(0))}{\dot{P}_{\rm c}(\rho_0(0))}\left[\frac{\ddot{P}_{\rm act}(\rho_0(0))}{\dot{P}_{\rm act}(\rho_0(0))} - \frac{\ddot{P}_{\rm c}(\rho_0(0))}{\dot{P}_{\rm c}(\rho_0(0))}\right] \rho_0'(0),
\end{equation}
where dots denote differentiation of the equations of state with respect to $\rho$. 
The value of the flat wall density profile at the wall is known exactly from $P_{\rm c}(\rho_0(0)) = P_{\rm c}(\rho_{\infty}) + P_{\rm act}(\rho_{\infty})$.
The value of its derivative at the wall is also known and revealed by the chain rule
\begin{equation}
    \rho_{0}'(0) = \frac{P'_c(0)}{\dot{P}_c(\rho_0(0))} = \frac{P_c^{(1)}}{\dot{P}_c(\rho_0(0))}
\end{equation}
where $P_c^{(1)}$ was found in the appendix of the main text. 

\subsection{\texorpdfstring{Approximating $\mathcal{K}(q)$}{Approximating K(q)}}
To approximate $\mathcal{K}(q) = \varphi'(\xi = 0)$ along the same lines, we need to work a little harder. 
First, recall that $\varphi(\xi)$ is the solution to
\begin{equation}
    \label{eq:appGreen}
	\Lambda_d^2 \varphi''(\xi) - Q(\mathbf{q},\xi) \varphi(\xi) = e^{-q \xi},
\end{equation}
subject to the boundary conditions $\varphi(0) = 0$ and $\varphi(\infty) = 0$.
This is solved formally with a Green function, $G_Q$, in the usual fashion
\begin{equation}
    \varphi(\xi) = \int_{0}^{\infty} d \xi' G_Q(\xi,\xi') e^{-q\xi'},
\end{equation}
where $G_Q$ solves 
\begin{equation}
    \label{eq:appGreen2}
	\Lambda_d^2 \frac{d^2}{d \xi^2}G_Q(\xi,\xi') - Q(\mathbf{q},\xi) G_Q(\xi,\xi') = \delta(\xi - \xi').
\end{equation}
This can also be analyzed using WKB theory~\cite{Bender1978} to obtain the standard expression
\begin{equation}
\label{eq:GQ}
    G_Q(\xi,\xi') = -\frac{1}{Q^{1/4}(\xi) Q^{1/4}(\xi')}\begin{cases}
e^{-\frac{1}{\Lambda_d} \int_{0}^{\xi'} d\zeta \sqrt{Q(\zeta)}} \sinh\left(\frac{1}{\Lambda_d} \int_{0}^{\xi} d\zeta \sqrt{Q(\zeta)}\right), \ \ \text{if} \ \ \xi < \xi'\\
 
\sinh\left(\frac{1}{\Lambda_d} \int_{0}^{\xi'} d\zeta \sqrt{Q(\zeta)}\right) e^{-\frac{1}{\Lambda_d} \int_{0}^{\xi} d\zeta \sqrt{Q(\zeta)}}  \ \ \text{if} \ \ \xi > \xi',
\end{cases}
\end{equation}
whence we obtain 
\begin{equation}
    \mathcal{K}(q) = -\frac{Q^{1/4}(0)}{\Lambda_d}\int_{0}^{\infty} d\xi \ e^{-q \xi}\frac{e^{-\frac{1}{\Lambda_d}\int_{0}^{\xi} d\zeta \sqrt{Q(\zeta)}}}{Q^{1/4}(\xi)}. 
\end{equation}
We cannot compute this integral exactly. However, we can approximate it by noting that the exponential factors in the integrand will strongly suppress any contributions away from $\xi = 0$. Therefore, we can expand the terms in the integrand as 
\begin{subequations}
    \begin{equation}
        Q^{-1/4}(\xi) \approx Q^{-1/4}(0) - \frac{Q'(0)}{4Q^{5/4}(0)}\xi + \mathcal{O}(\xi^2)
    \end{equation}
    and
    \begin{equation}
        \int_{0}^{\xi} d\zeta \sqrt{Q(\zeta)} \approx \sqrt{Q(0)}\xi + \frac{Q'(0)}{4\sqrt{Q(0)}} \xi^2 + \mathcal{O}(\xi^3).
    \end{equation}
\end{subequations}
This lets us obtain the approximation 
\begin{equation}
\label{eq:Kapprox}
    \mathcal{K}(q) \approx -\frac{1}{\Lambda_d q+\sqrt{Q(0)}} + \Lambda_d\frac{Q'(0)}{4 Q(0)} \frac{1}{(\Lambda_d q+\sqrt{Q(0)})^2} + \cdots.
\end{equation}
These are the first of a series of approximations to $\mathcal{K}(q)$, with each subsequent term including higher and higher derivatives or higher and higher powers of derivatives of $Q$ at $\xi=0$. Notice that this structure is \textit{the same} as arises  from the WKB approximation for $\psi(\xi)$ which leads to the approximation~\eqref{eq:Japprox} for $\mathcal{J}(q^2)$~\cite{Bender1978, Dingle1973}. Therefore, to the same order of accuracy of the WKB approximation, we need only keep the first two terms in the approximation for $\mathcal{K}(q)$. 

\subsection{Final expressions}
Using the results above, it is tedious but straightforward to obtain: 
\begin{equation}
    X_1/\Lambda_d = P_{\rm act}(0) - P_{\rm c}^{(1)} \mathcal{J}(0)
\end{equation}
\begin{equation}
    X_2/\Lambda_d^2 =P_{\rm c}^{(1)}  \left[1 + \mathcal{J}(0)\left(1- \mathcal{J}(0) + \mathcal{K}(0)\right)\right]+ P_{\rm act}(0) \ \left[\mathcal{J}(0) - \mathcal{K}(0) - 1\right]
\end{equation}
\begin{equation}
\begin{split}
X_3/\Lambda_d^3 &= P_{\rm c}^{(1)}  \left[
    - \mathcal{J}(0) ^ 3 
    + 2  \mathcal{J}(0)^ 2  ( 1 + \mathcal{K}(0)) 
    - \Lambda_d \mathcal{J}_1(0) 
    - \mathcal{J}(0)  (\mathcal{K}(0) ^ 2 + \mathcal{K}(0) - 1 - \Lambda_d \mathcal{K}_1(0))
    \right] \\
&+ P_{\rm act}(0)  \left[
        \mathcal{J}(0)^2
        + \mathcal{K}(0) 
        + \mathcal{K}(0)^2 
        - 2  \mathcal{J}(0)(1 + \mathcal{K}(0)) 
        - \Lambda_d \mathcal{K}_1(0)
    \right]
\end{split}
\end{equation}
\begin{equation}
\begin{split}
X_4/\Lambda_d^4 = & \frac{1}{2}\Bigg(
P_{\rm act}(0)\Big(
1
+ 2\,\mathcal{J}(0)^3
- 2\,\mathcal{K}(0)^2
- 2\,\mathcal{K}(0)^3
- 6\,\mathcal{J}(0)^2\bigl(1+\mathcal{K}(0)\bigr)
+ 2\,\Lambda_d \mathcal{J}_1(0) \\
&+ \mathcal{J}(0)\bigl(
2 + 8\,\mathcal{K}(0) + 6\,\mathcal{K}(0)^2 - 4\,\Lambda_d \mathcal{K}_1(0)
\bigr)
+ 2\,\Lambda_d \mathcal{K}_1(0)
+ 4\,\mathcal{K}(0)\Lambda_d \mathcal{K}_1(0)
- \Lambda_d^2 \mathcal{K}_2(0)
\Big) \\
& + P_{\rm c}^{(1)}\Big(
-2
- 2\,\mathcal{J}(0)^4
+ 6\,\mathcal{J}(0)^3\bigl(1+\mathcal{K}(0)\bigr)
+ 2\bigl(1+\mathcal{K}(0)\bigr)\Lambda_d \mathcal{J}_1(0)
+ \mathcal{J}(0)^2\bigl(
-8\,\mathcal{K}(0) - 6\,\mathcal{K}(0)^2 + 4\,\Lambda_d \mathcal{K}_1(0)
\bigr)\\
&
+ \mathcal{J}(0)\Big(
-3
+ 2\,\mathcal{K}(0)^2
+ 2\,\mathcal{K}(0)^3
- 4\,\Lambda_d \mathcal{J}_1(0)
- 2\,\Lambda_d \mathcal{K}_1(0)
- 2\,\mathcal{K}(0)\bigl(1+2\,\Lambda_d \mathcal{K}_1(0)\bigr)
+ \Lambda_d^2 \mathcal{K}_2(0)
\Big)
\Big)
\Bigg)
\end{split}
\end{equation}
In all of these expressions, subscripts denote differentiation with respect to $q$, i.e. $\mathcal{J}_n(0) = \lim_{q \to 0} d^n\mathcal{J}(q)/dq^n$. 
\subsection{Equations of State}
Equations~\eqref{eq:Japprox} and~\eqref{eq:Kapprox} allow each of these coefficients to be computed from the quantities, $\rho_0(0)$ and $P_{\rm c}^{(1)}$ which are determined by the equations of state $P_{\rm c}(\rho)$ and $P_{\rm act}(\rho)$ as well as the bulk density, $\rho_{\infty}$ and activity $\alpha$. The equations of state appropriate for hard-sphere ABPs in three dimensions used to produce Fig. 1 in the main text are provided below:
\begin{subequations}
    \begin{equation}
\frac{V_3(b)}{k_BT}P_c(\rho) 
=
\phi
+
\frac{
3\,\alpha\,\phi^2
}{
2^{2/3}\sqrt{1-\phi/\phi_{\rm max}}
}
    \end{equation}
    \begin{equation}
\frac{V_3(b)}{k_BT}P_{\mathrm{act}}(\rho)
=
\alpha^2 \phi\left[
1
+
\left(1-e^{-2^{5/3}\alpha}\right)\frac{
\phi
}{
1-\phi/\phi_{\rm max}
}\right]^{-1}.
    \end{equation}
\end{subequations}
Here $V_d(b)$ is the volume of a $d$-dimensional spherical ABP with radius $b$, i.e. $V_3(b) = 4 \pi b^3/3$, $\phi$ is the ABP volume fraction, $\phi = \rho V_d(b)$, the activity is $\alpha$, and $\phi_{\rm max} = 0.645$ is the maximum volume fraction attained in simulation~\cite{Omar2023}.
 
\section{Origins of the effective non-local terms}

In this section we discuss the origins of the $X_1$ and $X_3$ terms in the pressure exerted by the active particles on the membrane. 
These encode effective \textit{non-local} interactions between different points on the membrane surface and are mediated by the rearrangements of the particles above. 
Our goal here is not to present a detailed derivation of these terms (which would essentially recapitulate the calculation in the main text) but rather provide a physical picture that shows how they arise naturally from the governing equations. 
As our main focus will be their functional forms, unimportant constants and signs shall generally be ignored in this section, and the symbol ``$\sim$'' used to mean ``looks like'' or ``includes a term like''. 
We begin with a general heuristic that shows the origin of the $X_1 q$ term (recall $q = |\mathbf{q}|$ where $\mathbf{q}$ is the wave vector of a disturbance to the flat membrane) before explaining how this should be modified in the \textit{non-interacting case} where this term is \textit{absent} and the first non-local term is $X_3 q^3$.  

\subsection{General Argument}

As discussed in the main text, we are aiming to solve: 
\begin{equation}
\label{eq:governeq}
    \nabla^2 P_{\rm c} = \bm{\nabla} \cdot \bm{\mathcal{F}}, \ \ \ \text{and} \ \ \ \nabla^2 \bm{\mathcal{F}} - \bm{\mathcal{F}}= \bm{\nabla} P_{\rm act},
\end{equation}
for notational simplicity, here and henceforth, we non-dimensionalize lengths by the diffusion length, $\Lambda_d$, and work with reduced polarization $\bm{\mathcal{F}} = \Lambda_d \mathbf{F}$. 
Equations~\eqref{eq:governeq} are coupled non-linear partial differential equations for the density, $\rho$, via $P_{\rm c}$ and $P_{\rm act}$ and the polarization, via $\bm{\mathcal{F}}$. 

It is well known that active particles will preferentially orient perpendicular to a hard wall~\cite{Yan2015, Yan2018, Duzgun2018, Saintillan2015}. 
Therefore, it is natural to approximate $\bm{\mathcal{F}} \sim \hat{\mathbf{n}}(\mathbf{x}) \delta(z) $, where $\hat{\mathbf{n}}$ is the local normal to the membrane surface, $\mathbf{x}$ are the coordinates along it and $z$ is the normal coordinate to the membrane. 
We have taken the $z$ dependence of the polarization to be $\delta(z)$ to represent its, relatively rapid, decay to a homogeneous apolar state far from the membrane. 
The exact choice is not important for the argument, which is identical even when more realistic forms are used (e.g. exponential decay $\sim e^{- z/\ell_{z}}$), this one keeps our analysis as simple as possible. 
Using our approximation for $\bm{\mathcal{F}}$ in the first equation above, we have to solve
\begin{equation}
\label{eq:LR1}
    \nabla^2 P_{\rm c} \sim \delta(z) \ \bm{\nabla} \cdot \hat{\mathbf{n}}(\mathbf{x})
\end{equation}
and find Fourier transform (FT) of the pressure at the membrane, $\tilde{P}_{\rm c}(\mathbf{q} , z = 0)$.
Taking the full FT of both sides above, i.e. $ \tilde{f}(\mathbf{q},k) = \int d\mathbf{x} dz e^{i \mathbf{q} \cdot \mathbf{x}} e^{i k z} f(\mathbf{x},z)$, we find
\begin{equation}
   \tilde{P}_{\rm c}(\mathbf{q} ,k) \sim  \frac{q^2 \tilde{h}(q)}{k^2 + q^2}
\end{equation}
where we have expanded the normal for small membrane deviations, $\hat{\mathbf{n}} \sim \hat{\mathbf{z}} - \bm{\nabla} h$ and whence the value at $z = 0$ follows
\begin{equation}
    \tilde{P}_{\rm c}(\mathbf{q} , z = 0) \sim \int_{-\infty}^{\infty} dk \ \frac{q^2 \tilde{h}(\mathbf{q})}{k^2 + q^2} \sim q \ h(q),
\end{equation}
the integral in the last step can be evaluated using elementary methods (e.g. the residue theorem). 

This reveals how effective long-range interactions along the membrane arise. A local deformation of the membrane reorients nearby active particles (as captured by the right-hand side of Eq.~\eqref{eq:LR1}). The resulting particle currents redistribute the particle density above and along the membrane, through the Laplacian term on the left-hand side of Eq.~\eqref{eq:LR1}. This change in density in turn changes the pressure. As a consequence, changing the membrane curvature at one point affects the forces acting on the membrane at distant locations: the active particle suspension transmits this influence, effectively coupling remote regions of the membrane.

\subsection{Non-interacting case}

Having seen the general origin of a term $\sim X_1 q \tilde{h}$ in the pressure on the membrane, let us consider why this term is not present for non-interacting active particles. 
To do this, we shall use the mathematical construction of Yan and Brady~\cite{Yan2018} who considered an ideal active Brownian particle suspension outside a generally curved surface using boundary layer theory. 
In the absence of interactions, Eqs.~\eqref{eq:governeq} are simplified because $P_{\rm act} = \alpha^2 P_{\rm c}$. 
Upon defining $\bm{\nabla}\cdot \bm{\mathcal{F}} = \mathcal{G}$, it may be shown that, in this case, 
\begin{equation}
\label{eq:LR2}
    \nabla^2 \mathcal{G} - (1 + \alpha^2) \mathcal{G} = 0.
\end{equation}
We can use this to substitute for $\mathcal{G}$ in the first of Eq.~\eqref{eq:governeq} 
\begin{equation}
    \nabla^2 P_{\rm c} \sim \nabla^2 \mathcal{G} = \nabla^2 (\bm{\nabla} \cdot \bm{\mathcal{F}}).
\end{equation}
Taking the same form for $\bm{\mathcal{F}}$ and applying the same argument as in the previous section, we find $\tilde{P}_{\rm c}(\mathbf{q},z=0) \sim q^3 \tilde{h}$.
To get this term and not a term linear in $q$ it was necessary to use Eq.~\eqref{eq:LR2} which \textit{only holds in the non-interacting case} when $P_{\rm c} \propto P_{\rm act}$. 
Without this equation, there will always be a term that behaves $\sim \bm{\nabla} \cdot \bm{\mathcal{F}}$ on the right hand side of the first of Eqs.~\eqref{eq:governeq}. 

\section{Changing Membrane Properties}
Here, we show how changing the the surface tension, $\gamma$, and bending modulus, $\kappa$, alters the stability map as a function of particle volume fraction $\phi_{\infty}/\phi_{\rm max}$ and activity $\alpha$, shown in Fig.~1 of the main text. 
We use non-dimensionalized membrane parameters $g=\gamma V_d(b)/(\Lambda_d k_B T)$ and $k=\kappa V_d(b)/(\Lambda_d^3 k_B T)$, as in the main text.

\begin{figure}\includegraphics[width=16cm]{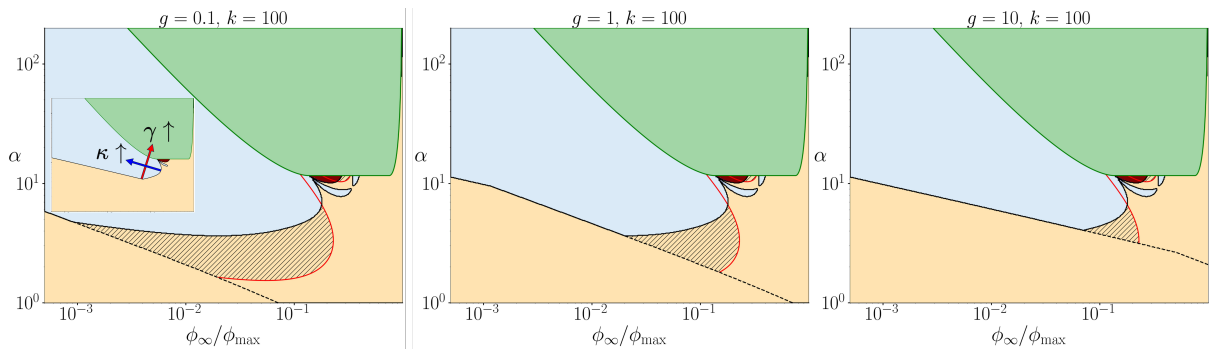}
	\caption{\label{fig:gsweep} 
Stability diagrams as a function of bulk volume fraction, $\phi_{\infty}/\phi_{\rm max}$, and activity, $\alpha$, for a fixed non-dimensional bare bending modulus $k=100$ as the non-dimensional tension is varied  from $g=0.1$ to $g=10$.
The inset shows the rough direction in which the unstable region shrinks as the membrane properties change.}
\end{figure}

\begin{figure}\includegraphics[width=16cm]{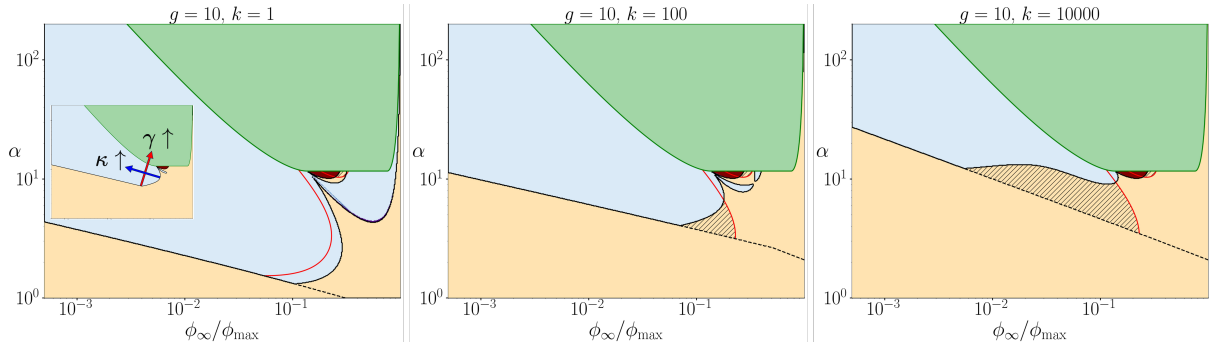}
	\caption{\label{fig:ksweep} 
Stability diagrams as a function of bulk volume fraction, $\phi_{\infty}/\phi_{\rm max}$, and activity, $\alpha$, for a fixed non-dimensional bare surface tension $g=10$ as the non-dimensional bending modulus is varied  from $k=1$ to $k=10,000$.
The inset shows the rough direction in which the unstable region shrinks as the membrane properties change.}
\end{figure}

For reference, in the simulations of Vutukuri et al.~\cite{Vutukuri2020} referenced in the main text, these parameters take values: $g=6$ and $k=154$. 
This motivates us to choose a representative value of $k=100$ and sweep $g$ from $0.1$ to $10$ in Fig.~\ref{fig:gsweep}.
This shows that the blue region (short wavelength instability) shrinks as surface tension increase. 
The rough direction in which it shrinks is indicated by the red arrow in the inset.
In Fig.~\ref{fig:ksweep}, we keep $g=10$ fixed and change $k$ from $1$ to $10,000$. 
This shows that the short-wavelength-unstable region shrinks as $k$ increases. 
Roughly speaking, the region shrinks in the direction perpendicular to that when $g$ increases, as indicated by the red arrow in the inset. 

\section{Details of Motility Induced Phase Separation Spinodal}

To determine where motility induced phase separation (MIPS) occurs, we use the spinodal criterion, ${\dot{\mathcal{P}}(\rho) = \dot{P}_{\rm c}(\rho) + \dot{P}_{\rm act}(\rho)} \leq 0$ (dots denote $\rho$ derivatives), that determines the linear stability of a homogeneous suspension to infinitessimal deviations.
Finite perturbations, e.g nucleation events, can destabilize a suspension \textit{between} the spinodal and the binodal, but we shall ignore these for simplicity. 

It is not immediately obvious how to apply the spinodal criterion to the stability of an interacting suspension of active particles outside a flat wall because the density profile, $\rho_0$, is not homogeneous. 
However, recall that if the function $Q(\mathbf{q},\xi) = (1+\Lambda_d^2 q^2 + \chi(\xi)) < 0$, then the perturbations to this density profile caused by membrane deformations will become oscillatory [see Eq.~\eqref{eq:fWKB}]. 
Using the definition of $\chi(\xi) = \dot{P}{\rm act}(\rho_0(\xi))/\dot{P}{\rm c}(\rho_0(\xi))$, where $\xi$ is the distance from the membrane, the stability of the passive suspension and the positivity of $\Lambda_d^2 q^2$, it follows that $Q>0$ at a point if $\dot{\mathcal{P}}(\rho_0(\xi)) > 0$ there too.
This immediately tells us that the bulk density $\rho_{\infty}$ must be outside the spinodal, i.e. $\dot{\mathcal{P}}(\rho_{\infty}) > 0$, because if the perturbations to the flat wall density oscillate infinitely far from the membrane, then they are incompatible with a homogeneous, apolar bulk suspension. 
Similarly, if the flat wall profile \textit{ever} reaches a density where $\dot{\mathcal{P}} < 0$, then our stability analysis will not hold because the osillatory corrections would need to be delicately matched to the exponentially decaying corrections in bulk~\cite{Bender1978, Dingle1973}. 
This restriction implies that the flat wall density profile is strictly monotonically decreasing from the value $\rho_0(0)$ to $\rho_{\infty}$. 
Therefore, in order to focus only on regions where MIPS does not occur (without activated nucleation events) we restrict our attention to desntities (and activities) where: $\dot{\mathcal{P}}(\rho) > 0,  \ \forall \rho \in [\rho_{\infty}, \rho_0(0)]$. 

\begin{figure}\includegraphics[width=8.8cm]{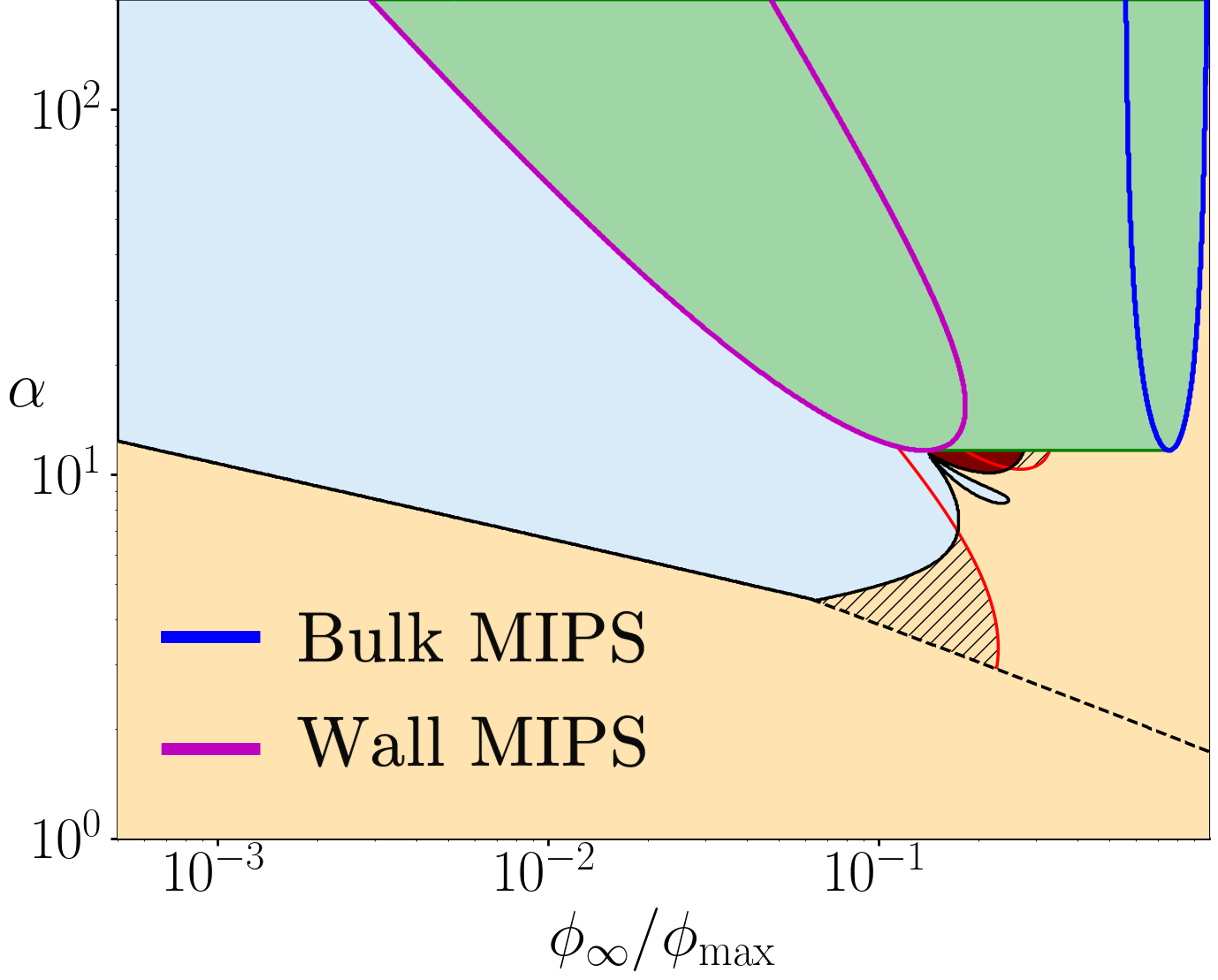}
	\caption{\label{fig:spinodal} 
Stability diagrams as a function of bulk volume fraction, $\phi_{\infty}/\phi_{\rm max}$, and activity, $\alpha$, for the membrane parameters used in the main text $g = 6$ and $k= 154$.
The spinodals corresponding the bulk and wall densities are shown in blue and pink respectively.
Given our assumptions, the flat wall density profile will reach all densities between these two values, and the composition of all of the corresponding spinodals gives the total MIPS region in green.}
\end{figure}

In Fig.~\ref{fig:spinodal} we show how this gives rise to the extended green MIPS region in our stability map (Fig. 1 of the main text). 
The extremal spinodal boundaries are shown: $\dot{\mathcal{P}}(\rho_{0}(0)) = 0$ in pink and $\dot{\mathcal{P}}(\rho_{\infty})$ in blue. 
The flat wall density profile at some point reaches every density between these two values, with each giving rise to a different spinodal. 
All of these regions are combined to give the total green MIPS region. 

%